\documentclass[preprint,showpacs,preprintnumbers,amsmath,amssymb,showkeys]{revtex4}

\usepackage{graphicx}% Include figure files
\usepackage{dcolumn}% Align table columns on decimal point
\usepackage{bm}% bold math
\usepackage{braket}
\usepackage{enumitem}
\usepackage[utf8]{inputenc}
\usepackage{amsmath}
\usepackage{amssymb}
\usepackage{amsthm}

\begin{document}

\noindent

\preprint{}

\title{Sufficient conditions, lower bounds and trade-off relations for quantumness in Kirkwood-Dirac quasiprobability}

\author{Agung Budiyono}
\email{agungbymlati@gmail.com}
\affiliation{Research Center for Quantum Physics, National Research and Innovation Agency, South Tangerang 15314, Republic of Indonesia}

\date{\today}% It is always \today, today,

\begin{abstract}

Kirkwood-Dirac (KD) quasiprobability is a quantum analog of classical phase space probability. It offers an informationally complete representation of quantum state wherein the quantumness associated with quantum noncommutativity manifests in its nonclassical values, i.e., the nonreal and/or negative values of the real part. This naturally raises a question: how does such form of quantumness comply with the uncertainty principle which also arise from quantum noncommutativity? Here, first, we obtain sufficient conditions for the KD quasiprobability defined relative to a pair of PVM (projection-valued measure) bases to have nonclassical values. Using these nonclassical values, we then introduce two quantities which capture the amount of KD quantumness in a quantum state relative to a single PVM basis. They are defined respectively as the nonreality, and the classicality which captures both the nonreality and negativity, of the associated KD quasiprobability over the PVM basis of interest, and another PVM basis, and maximized over all possible choices of the latter. We obtain their lower bounds, and derive trade-off relations respectively reminiscent of the Robertson and Robertson-Schr\"odinger uncertainty relations but with lower bounds maximized over the convex sets of Hermitian operators whose complete sets of eigenprojectors are given by the PVM bases. We discuss their measurement using weak value measurement and classical optimization. We then suggest an information theoretical interpretation of the KD nonreality relative to a PVM basis as a lower bound to the maximum total root-mean-squared error in an optimal estimation of the PVM basis, and thereby obtain a lower bound and a trade-off relation for the root-mean-squared error. Finally, we suggest an interpretation of the KD nonclassicality relative to a PVM basis as a lower bound to the total state disturbance caused by a nonselective projective binary measurement associated with the PVM basis, and derive a lower bound and a trade-off relation for the disturbance.  

\end{abstract} 

%Use showkeys class option if keyword 
\maketitle       

\section{Introduction}     

Heisenberg uncertainty principle is a basic tenet of quantum mechanics which sets down a radical conceptual demarcation from classical mechanics \cite{Heisenberg UR}. It stipulates a fundamental restriction, in the forms of trade-off relations, on the simultaneous predictability of outcomes of measurement of two physical quantities. Formally, the trade-off relations arise from the noncommutativity of operators representing quantum measurements \cite{Kennard UR,Weyl UR,Robertson UR}. From the very beginning, the uncertainty principle has led to the foundational debate about the deep nature of randomness arising in quantum measurement \cite{EPR paradox} and the intimately related conceptual issue on the meaning of quantum correlation \cite{Bell's theorem}. In recent decades, attempts to better understand the meaning of uncertainty relation and quantum randomness in general, has opened an avenue for fruitful applications in different areas of quantum science and quantum technology \cite{Coles entropic uncertainty relation review}. It is thus important to study the uncertainty principle from various perspectives to appreciate its rich and multi-faceted nature and to conceive further implications. 

The earliest uncertainty relations are developed based on the quantification of the measurement uncertainty in terms of variance of measurement outcomes \cite{Kennard UR,Weyl UR,Robertson UR,Schroedinger UR}. Certain drawbacks of variance for characterizing unpredictability motivated the construction of uncertainty relations based on the Shannon entropy of the measurement outcomes \cite{Hirschman UR based on entropy,Everett question on the UR based on entropy,Bialynicki-Birula UR based on entropy for phase space,Deutsch entropic UR,Maassen entropic UR,Berta entropic UR with side information,Coles entropic UR,Hall quantum-classical decomposition,Wehner entropic UR review,Coles entropic uncertainty relation review}. Variance and Shannon entropy of measurement outcomes however do not only quantify the genuine quantum uncertainty originating from the noncommutativity between the quantum state and the measurement operators. But, they also take into account the classical uncertainty stemming from the agent's ignorance about the preparation, either due to classical noise or lack of access of another system entangled with the system of interest, leading to the preparation of mixed states. It is thus instructive to ask if it is possible to develop uncertainty relations for the intrinsic quantum uncertainty rather than for the total measurement uncertainty. A notable result along this direction was reported in Ref. \cite{Luo quantum Robertson-like uncertainty relation for WY skew information} where the author derived a trade-off relation for an intrinsic quantum uncertainty quantified by means of Wigner-Yanase skew information \cite{Wigner-Yanase skew information}, having a form similar to the Robertson uncertainty relation. This result is generalized in Ref. \cite{Furuichi quantum Robertson-Schroedinger uncertainty relation for WY skew information} to obtain a trade-off relation similar to the Robertson-Schr\"odinger uncertainty relation. Another approach is suggested in Refs \cite{Korzekwa quantum-classical decomposition,Singh uncertainty relation for coherence,Yuan uncertainty relation for coherence,Hall quantum-classical decomposition} which used some measures of quantum coherence to isolate the intrinsic quantum uncertainty and showed that they satisfy some trade-off relations similar to the entropic uncertainty relations.   

In the present study, we work with an informationally equivalent representation of quantum states on a finite-dimensional Hilbert space using Kirkwood-Dirac (KD) quasiprobability \cite{Kirkwood quasiprobability,Dirac quasiprobability,Barut KD quasiprobability}. KD quasiprobability is a quantum analog of classical phase space probability wherein the quantumness associated with noncommutativity manifests in its nonclassical values, i.e., non-real values and/or negative values of its real part. This prompts the question on how the uncertainty principle imposes a restriction on such form of quantumness. In order to answer this question, we first derive sufficient conditions for the KD quasiprobability relative to a pair of rank-1 orthogonal PVM (projection-valued measure) bases to have nonclassical values. We then introduce two quantities which measure the KD quantumness in a quantum state relative to a single PVM basis. The first quantity is defined as the nonreality in the KD quasiprobability over the PVM basis of interest and another PVM basis, and maximized over all possible choices of the latter. We call it the KD nonreality in the quantum state relative the PVM basis. The second quantity is defined similarly, but relative to the nonclassicality which captures simultaneously both the nonreality and the negativity of the KD quasiprobability. We call it the KD nonclassicality in the quantum state relative the PVM basis. Both quantities have been proposed earlier in Refs. \cite{Agung KD-nonreality coherence,Agung KD-nonclassicality coherence} as faithful quantifiers of quantum coherence relative to the incoherent orthonormal basis corresponding to the rank-1 PVM basis. We obtain lower bounds for the quantumness captured by the above defined KD nonreality and KD nonclassicality in a state relative to a PVM basis. 

We then proceed to derive trade-off relations for the KD nonreality in a state relative to a PVM basis and that relative to another PVM basis, and similarly for the KD nonclassicality in a state relative to a PVM basis and that relative to another PVM basis. They are respectively reminiscent of the Robertson \cite{Robertson UR} and the Robertson-Schr\"odinger uncertainty relations \cite{Schroedinger UR}, but with lower bounds that are optimized over the convex sets of all pairs of Hermitian operators whose eigenprojectors are given by the two PVM bases of interest. The lower bounds and the trade-off relations for the KD nonreality and KD nonclassicality in a state relative to a rank-$1$ orthogonal PVM basis lead to similar lower bounds and trade-off relations for the $l_1$-norm coherence of the state relative to the incoherent orthonormal basis corresponding to the PVM basis \cite{Baumgratz quantum coherence measure}. We sketch a measurement scheme of the KD nonreality and KD nonclassicality relative to a PVM basis based on weak value measurement and classical optimization. We then suggest an information theoretical interpretation of the KD nonreality in a state relative to a PVM basis as a lower bound to the root-mean-squared error of an optimal estimation of the PVM basis based on projective measurement in the worst case scenario. This allows us to derive a lower bound and a trade-off relation for the root-mean-squared error of the optimal estimation of a PVM basis in the worst case scenario. We further suggest an operational interpretation of the KD nonclassicality in a state relative to a PVM basis as a lower bound to the total state disturbance caused by a nonselective projective binary measurement associated with the PVM basis, and thereby derive a lower bound and a trade-off relation of such state disturbance. 

\section{Sufficient conditions for nonclassical Kirkwood-Dirac quasiprobability}

KD quasiprobability is a specific quantum analog of phase space probability distribution in classical statistical mechanics \cite{Kirkwood quasiprobability,Dirac quasiprobability}. The KD quasiprobability associated with a quantum state represented by a density operator $\varrho$ on a Hilbert space $\mathcal{H}$ over a pair of orthonormal bases $\{\ket{a}\}$ and $\{\ket{b}\}$ of $\mathcal{H}$, is defined as \cite{Kirkwood quasiprobability,Dirac quasiprobability,Barut KD quasiprobability}
\begin{eqnarray}
{\rm Pr}_{\rm KD}(a,b|\varrho):={\rm Tr}\{\Pi_b\Pi_a\varrho\}, 
\label{standard KD quasiprobability}
\end{eqnarray}
where $\Pi_x:=\ket{x}\bra{x}$, $x=a,b$. We note that $\{\Pi_x\}$ comprises a set of rank-1 orthogonal PVM, i.e., $\sum_x\Pi_x=\mathbb{I}$, $\Pi_x\Pi_{x'}=\delta_{xx'}\Pi_x$, where $\mathbb{I}$ is the identity operator on $\mathcal{H}$ and $\delta_{xx'}$ is the Kronecker delta. The PVM $\{\Pi_x\}$ describes a sharp projective measurement with outcomes $x$ and probability ${\rm Pr}(x|\varrho)={\rm Tr}\{\Pi_x\varrho\}$. Here on we shall thus refer to $\{\Pi_x\}$ as a rank-1 PVM basis. 

KD quasiprobability gives correct marginal probabilities, i.e., $\sum_i{\rm Pr}_{\rm KD}(a,b|\varrho)={\rm Pr}(j|\varrho)$, $i\neq j$, $i,j=\{a,b\}$. However, unlike conventional classical probability, KD quasiprobability may take nonreal value, and its real part, called the Terletsky-Margenou-Hill quasiprobability \cite{Terletsky TBMH quasiprobability,Barut KD quasiprobability,Margenau TBMH quasiprobability}, may be negative. Such nonreality and negativity capture the quantum noncommutativity, that is, assuming two of its three ingredients $\{\varrho,\Pi_a,\Pi_b\}$ commute, e.g., $[\Pi_a,\varrho]_-=0$, renders the KD quasiprobability ${\rm Pr}_{\rm KD}(a,b|\varrho)$ real and nonnegative. Here and in what follows, $[X,Y]_{\mp}:=XY\mp YX$ denotes the commutator and anticommutator between two Hermitian operators $X$ and $Y$. In this sense, the nonreality or/and the negativity of KD quasiprobability delineate some form of quantumness stemming from quantum noncommutativity. The converse however is not necessarily true \cite{Drori nonclassicality tighter and noncommutativity,deBievre nonclassicality in KD distribution}. Remarkably, the real and imaginary parts of the KD quasiprobability can be estimated in experiment without resorting to full state tomography either using weak value measurement or other methods \cite{Aharonov weak value,Aharonov-Daniel book,Wiseman weak value,Lundeen measurement of KD distribution,Salvail direct measurement KD distribution,Bamber measurement of KD distribution,Thekkadath measurement of density matrix,Johansen quantum state from successive projective measurement,Lostaglio KD quasiprobability and quantum fluctuation,Hernandez-Gomez experimental observation of TBMH negativity,Wagner measuring weak values and KD quasiprobability,Vallone strong measurement to reconstruct quantum wave function,Cohen estimating of weak value with strong measurements,Lundeen complex weak value,Jozsa complex weak value}.  This form of quantumness, i.e., the nonreality or/and the negativity in the KD quasiprobability has thus found applications in different areas of quantum science and technology \cite{Lostaglio KD quasiprobability and quantum fluctuation,Allahverdyan TBMH as quasiprobability distribution of work,Lostaglio TBMH quasiprobability fluctuation theorem contextuality,Levy quasiprobability distribution for heat fluctuation in quantum regime,Alonso KD quasiprobability witnesses quantum scrambling,Halpern quasiprobability and information scrambling,Arvidsson-Shukur quantum advantage in postselected metrology,Lupu-Gladstein negativity enhanced quantum phase estimation 2022,Das KD quasiprobability in postselected metrology,Pusey negative TBMH quasiprobability and contextuality,Kunjwal contextuality of non-real weak value,Lostaglio contextuality in quantum linear response,Agung KD-nonreality coherence,Agung KD-nonclassicality coherence,Agung translational asymmetry from nonreal weak value,Agung estimation and operational interpretation of trace-norm asymmetry,Agung KD general quantum correlation}.

KD quasiprobability gives an informationally complete representation of an arbitrary quantum state. That is, given a KD quasiprobability ${\rm Pr}_{\rm KD}(a,b|\varrho)$ defined over a pair of orthonormal bases $\{\ket{a}\}$ and $\{\ket{b}\}$ with $\braket{a|b}\neq 0$ for all $(a,b)$, the associated quantum state can be reconstructed as $\varrho=\sum_{a,b}{\rm Pr}_{\rm KD}(a,b|\varrho)\frac{\ket{a}\bra{b}}{\braket{b|a}}$. This important fact naturally raises an intriguing question on how the KD quantumness capture different yet interrelated nonclassical concepts associated with a quantum state subjected to quantum measurements. To this end, we have argued previously that the nonreality or simultaneously both the nonreality and negativity of the KD quasiprobability can be used to quantitatively characterize quantum coherence \cite{Agung KD-nonreality coherence,Agung KD-nonclassicality coherence}, asymmetry \cite{Agung translational asymmetry from nonreal weak value,Agung estimation and operational interpretation of trace-norm asymmetry}, and general quantum correlation \cite{Agung KD general quantum correlation}. In the present article, we study how the quantumness in the KD quasiprobability complies with the quantum uncertainty principle. Both the KD quantumness and the uncertainty principle arise from the quantum noncommutativity. 

First, we summarize two mathematical objects for quantifying respectively the nonreality and the total nonclassicality which captures simultaneously both the nonreality and the negativity in the KD quasiprobability. To quantify the nonreality in the KD quasiprobability, we use the following $l_1$-norm of the nonreal part of the KD quasiprobability:
\begin{eqnarray}
{\rm NRe}(\{{\rm Pr}_{\rm KD}(a,b|\varrho)\})&:=&\sum_{a,b}|{\rm Im}{\rm Pr}(a,b|\varrho)|\nonumber\\
&=&\sum_{a,b}|{\rm Im}{\rm Tr}\{\Pi_b\Pi_a\varrho\}|. 
\label{KD nonreality}
\end{eqnarray} 
It vanishes if and only if the KD quasiprobability is real. Next, let us define the following quantity \cite{Drori nonclassicality tighter and noncommutativity,Alonso KD quasiprobability witnesses quantum scrambling,Lostaglio KD quasiprobability and quantum fluctuation}:
\begin{eqnarray}
{\rm NCl}(\{{\rm Pr}_{\rm KD}(a,b|\varrho)\})&:=&\sum_{a,b}|{\rm Pr}_{\rm KD}(a,b|\varrho)|-1\nonumber\\
&=&\sum_{a,b}|{\rm Tr}\{\Pi_b\Pi_a\varrho\}|-1.  
\label{KD nonclassicality}
\end{eqnarray}
It is nonnegative by definition since $\sum_{a,b}|{\rm Pr}_{\rm KD}(a,b|\varrho)|\ge |\sum_{a,b}{\rm Pr}_{\rm KD}(a,b|\varrho)|=1$, where the equality follows from the fact that KD quasiprobability is always normalized, i.e., $\sum_{a,b}{\rm Pr}_{\rm KD}(a,b|\varrho)=1$. Moreover, it vanishes only when $|{\rm Pr}_{\rm KD}(a,b|\varrho)|={\rm Pr}_{\rm KD}(a,b|\varrho)$ for all $a$ and $b$, i.e., only when ${\rm Pr}_{\rm KD}(a,b|\varrho)$ is real and nonnegative. ${\rm NCl}(\{{\rm Pr}_{\rm KD}(a,b|\varrho)\})$ defined in Eq (\ref{KD nonclassicality}) thus quantifies the failure of the KD quasiprobability ${\rm Pr}_{\rm KD}(a,b|\varrho)$ to be both real and nonnegative. We refer to ${\rm NRe}(\{{\rm Pr}_{\rm KD}(a,b|\varrho)\})$ and ${\rm NCl}(\{{\rm Pr}_{\rm KD}(a,b|\varrho)\})$ defined respectively in Eqs. (\ref{KD nonreality}) and (\ref{KD nonclassicality}) as the KD nonreality and the KD nonclassicality in the quantum state $\varrho$ relative to the pair of PVM bases $\{\Pi_a\}$ and $\{\Pi_b\}$.   

We obtain two simple sufficient conditions respectively for nonvanishing ${\rm NRe}(\{{\rm Pr}_{\rm KD}(a,b|\varrho)\})$ and ${\rm NCl}(\{{\rm Pr}_{\rm KD}(a,b|\varrho)\})$. Below, we use the notation $\|X\|_{\infty}$ to denote the operator norm or the $\infty$-Schatten norm of an operator $X$. $\|X\|_{\infty}$ is equal to the largest eigenvalue modulus or the spectral radius of $X$. Using the operator norm of a Hermitian operator $X$, we then define the corresponding normalized Hermitian operator as $\tilde{X}:=X/\|X\|_{\infty}$. 

First, we have the following result for the KD nonreality in a state relative to a pair of PVM bases. \\
{\bf Lemma 1}. Given a state $\varrho$ on a Hilbert space $\mathcal{H}$, the nonreality in the associated KD quasiprobability over a pair of PVM bases $\{\Pi_a\}$ and $\{\Pi_b\}$ of $\mathcal{H}$ defined in Eq. (\ref{KD nonreality}), is lower bounded as 
\begin{eqnarray}
{\rm NRe}(\{{\rm Pr}_{\rm KD}(a,b|\varrho)\})\ge\frac{1}{2}\big|{\rm Tr}\{\tilde{B}[\tilde{A},\varrho]_-\}\big|,
\label{KD nonreality relative to a pair of PVMs is lower bounded by Robertson bound}
\end{eqnarray}
where $A$ and $B$ are any Hermitian operators with bounded spectrum whose complete set of eigenprojectors are respectively given by $\{\Pi_a\}$ and $\{\Pi_b\}$. \\
{\bf Proof}. Let $A=\sum_aa\Pi_a$ be a Hermitian operator on $\mathcal{H}$ with the complete set of eigenprojectors $\{\Pi_{a}\}$ and the associated spectrum of eigenvalues $\{a\}$. Similarly, let $B=\sum_bb\Pi_b$ be a Hermitian operator on $\mathcal{H}$ with the complete set of eigenprojectors $\{\Pi_b\}$ and the associated spectrum of eigenvalues $\{b\}$. From the definition of the KD nonreality in the quantum state $\varrho$ relative to a pair of PVM bases $\{\Pi_a\}$ and $\{\Pi_b\}$ in Eq. (\ref{KD nonreality}), we have
\begin{eqnarray}
{\rm NRe}(\{{\rm Pr}_{\rm KD}(a,b|\varrho)\})&=&\frac{1}{\|A\|_{\infty}\|B\|_{\infty}}\sum_{a,b}\|A\|_{\infty}\|B\|_{\infty}\big|{\rm Im}({\rm Tr}\{\Pi_b\Pi_a\varrho\})\big|\nonumber\\
\label{KD nonreality is lower bounded by Robertson bound 1 step 1}
&\ge&\frac{1}{\|A\|_{\infty}\|B\|_{\infty}}\big|{\rm Im}{\rm Tr}\{BA\varrho\}\big|\\
\label{KD nonreality relative to a PVM is lower bounded by Robertson bound 1 step 2}
%&=&\frac{1}{2\|A\|_{\infty}\|B\|_{\infty}}\big|{\rm Tr}\{B[A,\varrho]_-\}\big|\nonumber\\
&=&\frac{1}{2}\big|{\rm Tr}\{\tilde{B}[\tilde{A},\varrho]_-\}\big|, \nonumber
\label{KD nonreality relative to a PVM is lower bounded by Robertson bound 1 step 3}
\end{eqnarray}
where the inequality in Eq. (\ref{KD nonreality is lower bounded by Robertson bound 1 step 1}) is due to the fact that $\|A\|_{\infty}=\max\{|a|\}$ and $\|B\|_{\infty}=\max\{|b|\}$ and triangle inequality. \qed 

As an immediate corollary of the Lemma 1, while noncommutativity of all pairs of $A$, $B$, and $\varrho$ are not sufficient for the KD quasiprobability ${\rm Pr}_{\rm KD}(a,b|\varrho)$ associated with $\varrho$ defined over the eigenbasis $\{\ket{a}\}$ of $A$ and the eigenbasis $\{\ket{b}\}$ of $B$ to have nonreal value (or its real part is negative, or both) \cite{Drori nonclassicality tighter and noncommutativity,deBievre nonclassicality in KD distribution,deBievre incompatibility-uncertainty-KD nonclassicality,Xu KD classical pure states}, a nonvanishing lower bound in Eq. (\ref{KD nonreality relative to a pair of PVMs is lower bounded by Robertson bound}), i.e., ${\rm Tr}\{B[A,\varrho]_-\}\neq 0$ for a pair of Hermitian operators $A$ and $B$, is sufficient for the corresponding KD quasiprobability ${\rm Pr}_{\rm KD}(a,b|\varrho)$ to be nonreal for some $(a,b)$. It is interesting to remark that the lower bound in Eq. (\ref{KD nonreality relative to a pair of PVMs is lower bounded by Robertson bound}) takes a form similar to that of the Robertson uncertainty relation \cite{Robertson UR}.

Next, we derive a lower bound for the KD nonclassicality in a state relative to a pair of PVM bases defined in Eq. (\ref{KD nonclassicality}). \\
{\bf Lemma 2}. Given a state $\varrho$ on a Hilbert space $\mathcal{H}$, the nonclassicality in the associated KD quasiprobability associated with $\varrho$ over a pair of PVM bases $\{\Pi_{a}\}$ and $\{\Pi_b\}$ of $\mathcal{H}$ defined in Eq. (\ref{KD nonclassicality}) is lower bounded as
\begin{eqnarray}
{\rm NCl}(\{{\rm Pr}_{\rm KD}(a,b|\varrho)\})&\ge&\frac{1}{2}\big(\big|{\rm Tr}\{\varrho[\tilde{A}_{\varrho},\tilde{B}_{\varrho}]_-\}\big|^2\nonumber\\
&+&\big|{\rm Tr}\{\varrho[\tilde{A}_{\varrho},\tilde{B}_{\varrho}]_+\}-2{\rm Tr}\{\tilde{A}_{\varrho}\varrho\}{\rm Tr}\{\tilde{B}_{\varrho}\varrho\}\big|^2\big)^{1/2}-1, 
\label{KD nonclassility relative to a pair of POVMs is lower bounded by Robertson-Schrodinger bound}
\end{eqnarray}
where $\tilde{X}_{\varrho}:=\frac{X}{\|X-{\rm Tr}\{X\varrho\}\mathbb{I}\|_{\infty}}$, $X=A,B$, and $A$ and $B$ are any Hermitian operators with bounded spectrum whose complete set of eigenprojectors are respectively given by $\{\Pi_a\}$ and $\{\Pi_b\}$. \\
{\bf Proof}. Let again $A=\sum_aa\Pi_a$ be a Hermitian operator on $\mathcal{H}$ with the complete set of eigenprojectors $\{\Pi_{a}\}$ and the associated spectrum of eigenvalues $\{a\}$. Likewise, let $B=\sum_bb\Pi_b$ be a Hermitian operator on $\mathcal{H}$ with the complete set of eigenprojectors $\{\Pi_b\}$ and the associated spectrum of eigenvalues $\{b\}$. Then, from the definition of the KD nonclassicality in $\varrho$ relative to a pair of PVM bases $\{\Pi_a\}$ and $\{\Pi_b\}$ in Eq. (\ref{KD nonclassicality}), we first have
\begin{eqnarray}
{\rm NCl}(\{{\rm Pr}_{\rm KD}(a,b|\varrho)\})&=&\sum_{a,b}\big|{\rm Tr}\{\Pi_b\Pi_a\varrho\}\big|-1\nonumber\\
&=&\frac{\sum_{a,b}\|A\|_{\infty}\|B\|_{\infty}\big|{\rm Tr}\{\Pi_b\Pi_a\varrho\}\big|}{\|A\|_{\infty}\|B\|_{\infty}}-1\nonumber\\
%&\ge&\frac{1}{\|A\|_{\infty}\|B\|_{\infty}}\big|{\rm Tr}\{\varrho BA\}\big|\nonumber\\
&\ge&\big|{\rm Tr}\{\varrho\tilde{B}\tilde{A}\}\big|-1\nonumber\\
&=&\frac{1}{2}\big|{\rm Tr}\{\varrho[\tilde{B},\tilde{A}]_-\}+{\rm Tr}\{\varrho[\tilde{A},\tilde{B}]_+\}\big|-1,
\label{KD nonreality relative to a pair of PVM is lower bounded by Robertson-Schroedinger bound 1 step 3}
\end{eqnarray}
where to get the last line, we have used a decomposition: $\tilde{B}\tilde{A}=\frac{1}{2}[\tilde{B},\tilde{A}]_-+\frac{1}{2}[\tilde{A},\tilde{B}]_+$. Notice that ${\rm Tr}\{\varrho[\tilde{B},\tilde{A}]_-\}$ is pure imaginary while ${\rm Tr}\{\varrho[\tilde{A},\tilde{B}]_+\}$ is real. Hence, the modulus in Eq. (\ref{KD nonreality relative to a pair of PVM is lower bounded by Robertson-Schroedinger bound 1 step 3}) can be evaluated to give
\begin{eqnarray}
&&{\rm NCl}(\{{\rm Pr}_{\rm KD}(a,b|\varrho)\})\ge\frac{1}{2}\big(\big|{\rm Tr}\{\varrho[\tilde{B},\tilde{A}]_-\}\big|^2+\big|{\rm Tr}\{\varrho[\tilde{A},\tilde{B}]_+\}\big|^2\big)^{1/2}-1. 
\label{KD nonreality relative to a pair of PVM is lower bounded by Robertson-Schroedinger bound 1 step 4}
\end{eqnarray}
Next, note that the left-hand side in Eq. (\ref{KD nonreality relative to a pair of PVM is lower bounded by Robertson-Schroedinger bound 1 step 4}) does not depend on the spectrum of eigenvalues of $A$ and $B$. Now, consider the following Hermitian operators $A'=\sum_a(a-{\rm Tr}\{A\varrho\})\Pi_a=A-{\rm Tr}\{A\varrho\}\mathbb{I}$ and $B'=\sum_b(b-{\rm Tr}\{B\varrho\})\Pi_b=B-{\rm Tr}\{B\varrho\}\mathbb{I}$. Then, we have ${\rm Tr}\{\varrho[A',B']_-\}={\rm Tr}\{\varrho[A,B]_-\}$ and ${\rm Tr}\{\varrho[A',B']_+\}={\rm Tr}\{\varrho[A,B]_+\}-2{\rm Tr}\{A\varrho\}{\rm Tr}\{B\varrho\}$. Using these relations, replacing $A$ and $B$ in Eq. (\ref{KD nonreality relative to a pair of PVM is lower bounded by Robertson-Schroedinger bound 1 step 4}) respectively with $A'$ and $B'$, we obtain Eq. (\ref{KD nonclassility relative to a pair of POVMs is lower bounded by Robertson-Schrodinger bound}). \qed

Lemma 2 shows that a nonvanishing lower bound in Eq. (\ref{KD nonclassility relative to a pair of POVMs is lower bounded by Robertson-Schrodinger bound}), i.e., $\frac{1}{2}\big(\big|{\rm Tr}\{\varrho[\tilde{A}_{\varrho},\tilde{B}_{\varrho}]_-\}\big|^2+\big|{\rm Tr}\{\varrho[\tilde{A}_{\varrho},\tilde{B}_{\varrho}]_+\}-2{\rm Tr}\{\tilde{A}_{\varrho}\varrho\}{\rm Tr}\{\tilde{B}_{\varrho}\varrho\}\big|^2\big)^{1/2}-1 > 0$, provides a sufficient condition for the associated KD quasiprobability ${\rm Pr}_{\rm KD}(a,b|\varrho)$ to be nonreal, or its real part is negative, or both, for some $(a,b)$. It is again interesting to note that the lower bound takes a form similar to the lower bound of the Robertson-Schr\"odinger uncertainty relation. Unlike the latter, however, the lower bound in Eq. (\ref{KD nonclassility relative to a pair of POVMs is lower bounded by Robertson-Schrodinger bound}) depends nonlinearly on the state. Note that the sufficient condition in Lemma 2 is stronger than that in Lemma 1 since the former can also detect negativity of the KD quasiprobability. 

\section{Lower bounds and trade-off relations for KD quantumness in a state relative to a single rank-1 orthogonal PVM basis}

We first stress that both ${\rm NRe}(\{{\rm Pr}_{\rm KD}(a,b|\varrho)\})$ and ${\rm NCl}(\{{\rm Pr}_{\rm KD}(a,b|\varrho)\})$ defined in Eqs. (\ref{KD nonreality}) and (\ref{KD nonclassicality}) quantify the KD quantumness stemming from the failure of commutativity between the state $\varrho$ and both of the rank-1 PVMs bases  $\{\Pi_a\}$ and $\{\Pi_{b}\}$, and also between the pair of the PVMs bases. How does the quantumness of the KD quasiprobability portray the noncommutativity between a state and a single PVM basis, e.g., between $\varrho$ and the PVM basis $\{\Pi_a\}$? Quantities which reliably capture the noncommutativity between a state $\varrho$ and a single PVM basis $\{\Pi_a\}$ is desirable as we discuss the relation between the quantumness of the KD quasiprobability and the uncertainty in measurement described by the rank-1 PVM $\{\Pi_a\}$ over the state $\varrho$, and the associated uncertainty relations. To this end we introduce the following two quantities. 
\\
{\bf Definition 1}. The KD nonreality in a state $\varrho$ on a finite-dimensional Hilbert space $\mathcal{H}$ relative to a PVM basis $\{\Pi_a\}$ of $\mathcal{H}$ is defined as 
\begin{eqnarray}
\mathcal{Q}_{\rm KD}^{\rm NRe}(\varrho;\{\Pi_{a}\})&:=&\sup_{\{\Pi_{b}\}\in\mathcal{M}_{\rm r1PVM}(\mathcal{H})}{\rm NRe}(\{{\rm Pr}_{\rm KD}(a,b|\varrho)\})\nonumber\\
&=&\sup_{\{\Pi_{b}\}\in\mathcal{M}_{\rm r1PVM}(\mathcal{H})}\sum_{a,b}\big|{\rm Im}{\rm Pr}_{\rm KD}(a,b|\varrho)\big|,
%&=&\sup_{\{\Pi_{b}\}\in\mathcal{M}_{\rm r1PVM}(\mathcal{H})}\sum_{a,b}|{\rm Im}{\rm Tr}\{\Pi_b\Pi_a\varrho\}|,
\label{KD nonreality relative to a PVM}
\end{eqnarray}
where the supremum is taken over the set $\mathcal{M}_{\rm r1PVM}(\mathcal{H})$ of all the rank-1 PVM bases of $\mathcal{H}$. 
\\
{\bf Definition 2}. The KD nonclassicality in a state $\varrho$ on a finite-dimensional Hilbert space $\mathcal{H}$ relative to a PVM $\{\Pi_a\}$ of $\mathcal{H}$ is defined as
\begin{eqnarray}
\mathcal{Q}_{\rm KD}^{\rm NCl}(\varrho;\{\Pi_{a}\})&:=&\sup_{\{\Pi_{b}\}\in\mathcal{M}_{\rm r1PVM}(\mathcal{H})}{\rm NCl}(\{{\rm Pr}_{\rm KD}(a,b|\varrho)\})\nonumber\\
&=&\sup_{\{\Pi_{b}\}\in\mathcal{M}_{\rm r1PVM}(\mathcal{H})}\sum_{a,b}|{\rm Tr}\{\Pi_b\Pi_a\varrho\}|-1. 
\label{KD nonclassicality relative to a PVM}
\end{eqnarray}

Let us mention that $\mathcal{Q}_{\rm KD}^{\rm NRe}(\varrho;\{\Pi_{a}\})$ and $\mathcal{Q}_{\rm KD}^{\rm NCl}(\varrho;\{\Pi_{a}\})$ defined respectively in Eqs. (\ref{KD nonreality relative to a PVM}) and (\ref{KD nonclassicality relative to a PVM}) have been introduced earlier in Refs.  \cite{Agung KD-nonreality coherence,Agung KD-nonclassicality coherence}. There, it is argued that both quantities can be used as faithful quantifiers of coherence of $\varrho$ relative to the incoherent orthonormal basis $\{\ket{a}\}$ corresponding to the rank-1 orthogonal PVM basis $\{\Pi_a\}$ possessing certain desirable properties. In particular, one can show that both quantities are vanishing if and only if the state and the measurement basis are commuting: $[\Pi_a,\varrho]_-=0$ for all $a$ so that $\varrho$ is incoherent relative to the orthonormal basis $\{\ket{a}\}$.

In the following subsections, we will derive lower bounds and trade-off relations for the KD nonreality and KD nonclassicality in a quantum state relative to a PVM basis defined respectively in Eqs. (\ref{KD nonreality relative to a PVM}) and (\ref{KD nonclassicality relative to a PVM}). For this purpose, we denote by $\mathbb{H}(\mathcal{H})$ the convex set of all bounded Hermitian operators on the Hilbert space $\mathcal{H}$, $\mathbb{H}(\mathcal{H}|\{\Pi_x\})$ is the convex set of all bounded Hermitian operators on $\mathcal{H}$ having the complete set of eigenprojectors $\{\Pi_x\}$, and $\mathbb{H}(\mathcal{H}|\{x\})$ denotes the set of all bounded Hermitian operators with a spectrum of eigenvalues $\{x\}$, $x\in\mathbb{R}$. 

\subsection{Lower bound and trade-off relation for the KD nonreality in a state relative to a PVM basis} 

Using Lemma 1, we directly obtain a lower bound for the quantumness associated with the KD nonreality in a quantum state relative to a PVM basis. \\
{\bf Proposition 1}. The KD nonreality in a state $\varrho$ on a finite-dimensional Hilbert space $\mathcal{H}$ relative to a PVM basis $\{\Pi_{a}\}$ of $\mathcal{H}$ defined in Eq. (\ref{KD nonreality relative to a PVM}) is lower bounded as
\begin{eqnarray}
\label{KD nonreality relative to a PVM is lower bounded by a normalized Robertson bound}
&&\mathcal{Q}_{\rm KD}^{\rm NRe}(\varrho;\{\Pi_{a}\})\ge\frac{1}{2}\sup_{A\in\mathbb{H}(\mathcal{H}|\{\Pi_{a}\})}\sup_{B\in\mathbb{H}(\mathcal{H})}\big|{\rm Tr}\{\tilde{B}[\tilde{A},\varrho]_-\}\big|.
\label{KD nonreality relative to a PVM is lower bounded by a normalized noncommutativity}
\end{eqnarray}
\\
{\bf Proof}. Taking the supremum over the set $\mathcal{M}_{\rm r1PVM}(\mathcal{H})$ of all the rank-1 PVM bases $\{\Pi_b\}$ of $\mathcal{H}$ to both sides of Eq. (\ref{KD nonreality relative to a pair of PVMs is lower bounded by Robertson bound}), and noting Eq. (\ref{KD nonreality relative to a PVM}), we first have 
\begin{eqnarray}
\mathcal{Q}_{\rm KD}^{\rm NRe}(\varrho;\{\Pi_{a}\})&=&\sup_{\{\Pi_{b}\}\in\mathcal{M}_{\rm r1PVM}(\mathcal{H})}{\rm NRe}(\{{\rm Pr}_{\rm KD}(a,b|\varrho)\})\nonumber\\
&\ge&\frac{1}{2}\sup_{B\in\mathbb{H}(\mathcal{H}|\{b\})}\big|{\rm Tr}\{\tilde{B}[\tilde{A},\varrho]_-\}\big|.
\label{KD nonreality relative to a PVM is lower bounded by Robertson bound proof step 1}
\end{eqnarray} \\
Next, notice that the left-hand side of Eq. (\ref{KD nonreality relative to a PVM is lower bounded by Robertson bound proof step 1}) depends only on the PVM basis $\{\Pi_a\}$, i.e., it is independent of the spectrum of eigenvalues $\{a\}$ of $A$ and the spectrum of eigenvalues $\{b\}$ of $B$. Hence, upon further taking the supremum over all possible eigenvalues spectrum of $A$ and that of $B$ on the right-hand side of Eq. (\ref{KD nonreality relative to a PVM is lower bounded by Robertson bound proof step 1}), the inequality can be strengthened as in Eq. (\ref{KD nonreality relative to a PVM is lower bounded by a normalized Robertson bound}). 
\qed

%A couple of notes are in order. Assume that the state $\varrho$ is commuting with the PVM basis $\{\Pi_a\}$ for all $a$ so that $\mathcal{Q}_{\rm KD}^{\rm NRe}(\varrho;\{\Pi_{a}\})=0$. In this case, we also have $[A,\varrho]_-=0$ for all $A\in\mathbb{H}(\mathcal{H}|\{\Pi_a\})$. Hence, as expected, the lower bound in Eq. (\ref{KD nonreality relative to a PVM is lower bounded by a normalized Robertson bound}) is also vanishing. In particular, for a maximally mixed state, $\varrho=\mathbb{I}/d$, where $\mathbb{I}$ is an identity operator and $d$ is the dimension of the Hilbert space, both sides of the inequality in Eq. (\ref{KD nonreality relative to a PVM is lower bounded by a normalized Robertson bound}) are vanishing. 

The lower bound in Eq. (\ref{KD nonreality relative to a PVM is lower bounded by a normalized Robertson bound}) can be further evaluated to have the following result. \\
{\bf Proposition 2}. The KD nonreality in a quantum state $\varrho$ on a finite-dimensional Hilbert space $\mathcal{H}$ relative to a PVM basis $\{\Pi_{a}\}$ of $\mathcal{H}$ is lower bounded by the maximum trace-norm asymmetry of the state relative to the translation group generated by all Hermitian operators with the complete set of eigenprojectors that is given by $\{\Pi_{a}\}$ as
\begin{eqnarray}
\mathcal{Q}_{\rm KD}^{\rm NRe}(\varrho;\{\Pi_{a}\})\ge\sup_{A\in\mathbb{H}(\mathcal{H}|\{\Pi_{a}\})}\|[A,\varrho]_-\|_1/2\|A\|_{\infty}.
\label{KD nonreality is lower bounded by a normalized trace-norm asymmetry}
\end{eqnarray}
Here, $\|O\|_1={\rm Tr}\{\sqrt{OO^{\dagger}}\}$ is the Schatten 1-norm or the trace-norm of operator $O$, and $\|[A,\varrho]_-\|_1/2$ is just the trace-norm asymmetry of the state $\varrho$ relative to the group of translation unitary generated by the Hermitian operator $A$ \cite{Marvian - Spekkens speakable and unspeakable coherence}. \\
{\bf Proof}. See Appendix \ref{A proof of Proposition 2}. 

We show in Appendix \ref{Proof that the equality in lemma 1 is achieved for a single qubit} that for two-dimensional Hilbert space $\mathbb{H}\cong\mathbb{C}^2$, i.e., a system of a single qubit, both the inequalities in Eqs. (\ref{KD nonreality relative to a PVM is lower bounded by a normalized Robertson bound}) and (\ref{KD nonreality is lower bounded by a normalized trace-norm asymmetry}) become equalities for arbitrary state $\varrho$ on $\mathbb{C}^2$ and arbitrary PVM basis $\{\Pi_a\}$ of $\mathbb{C}^2$. In this case, both sides in  Eqs. (\ref{KD nonreality relative to a PVM is lower bounded by a normalized Robertson bound}) and (\ref{KD nonreality is lower bounded by a normalized trace-norm asymmetry}) are given by the corresponding $l_1$-norm coherence of a state $\varrho$ relative to the incoherent orthonormal basis $\{\ket{a}\}$ defined as $C_{l_1}(\varrho;\{\ket{a}\}):=\sum_{a\neq a'}|\braket{a|\varrho|a'}|$ \cite{Baumgratz quantum coherence measure}, directly quantifying the total magnitude of the off-diagonal terms of the density matrix. Hence, for any state $\varrho$ on $\mathbb{C}^2$ and any PVM basis $\mathbb{A}=\{\ket{e}\bra{e},\ket{e_{\perp}}\bra{e_{\perp}}\}$ of $\mathbb{C}^2$, where $\ket{e_{\perp}}$ is the orthonormal partner of $\ket{e}$, we have 
\begin{eqnarray}
\mathcal{Q}_{\rm KD}^{\rm NRe}(\varrho;\mathbb{A})&=&\frac{1}{2}\sup_{A\in\mathbb{H}(\mathbb{C}^2|\mathbb{A})}\sup_{B\in\mathbb{H}(\mathbb{C}^2)}\big|{\rm Tr}\{\tilde{B}[\tilde{A},\varrho]_-\}\big|\nonumber\\
&=&\sup_{A\in\mathbb{H}(\mathbb{C}^2|\mathbb{A})}\|[\tilde{A},\varrho]_-\|_1/2\nonumber\\
&=&2|\braket{e|\varrho|e_{\perp}}|\nonumber\\
&=&C_{l_1}(\varrho;\{\ket{e},\ket{e_{\perp}}\}). 
\label{KD-nonreality for two-dimensional Hilbert space}
\end{eqnarray}
Moreover, the eigenbasis of $B_*$, where $B_*\in\mathbb{H}(\mathbb{C}^2)$ is a Hermitian operator which attains the supremum in Eq. (\ref{KD-nonreality for two-dimensional Hilbert space}), is mutually unbiased with the orthonormal reference basis $\{\ket{e},\ket{e_{\perp}}\}$ and also with the eigenbasis of $\varrho$.

We finally obtain the following trade-off relation. \\
{\bf Proposition 3}. The KD nonreality in a quantum state $\varrho$ on a finite-dimensional Hilbert space $\mathcal{H}$ relative to a rank-1 PVM basis $\{\Pi_{a}\}$ of $\mathcal{H}$ and that relative to another rank-1 PVM basis $\{\Pi_{b}\}$ of $\mathcal{H}$ satisfy the following trade-off relation:
\begin{eqnarray}
&&\mathcal{Q}_{\rm KD}^{\rm NRe}(\varrho;\{\Pi_{a}\})\mathcal{Q}_{\rm KD}^{\rm NRe}(\varrho;\{\Pi_{b}\})\ge\frac{1}{4}\sup_{A\in\mathbb{H}(\mathcal{H}|\{\Pi_{a}\})}\sup_{B\in\mathbb{H}(\mathcal{H}|\{\Pi_{b}\})}\big|{\rm Tr}\{[\tilde{A},\tilde{B}]_-\varrho\}\big|^2. 
\label{uncertainty relation for the KD nonreality relative to a PVM}
\end{eqnarray} 
\\
{\bf Proof}. We first write the inequality in Eq. (\ref{KD nonreality relative to a PVM is lower bounded by a normalized Robertson bound}) as 
\begin{eqnarray}
&&\mathcal{Q}_{\rm KD}^{\rm NRe}(\varrho;\{\Pi_{a}\})\ge\frac{1}{2}\sup_{A\in\mathbb{H}(\mathcal{H}|\{\Pi_{a}\})}\sup_{B\in\mathbb{H}(\mathcal{H})}\big|{\rm Tr}\{[\tilde{A},\tilde{B}]_-\varrho\}\big|. 
\label{uncertainty relation for the KD nonreality relative to a PVM proof step 1}
\end{eqnarray}
Next, exchanging the role of $A$ and $B$ in Eq. (\ref{uncertainty relation for the KD nonreality relative to a PVM proof step 1}), we also have 
\begin{eqnarray}
&&\mathcal{Q}_{\rm KD}^{\rm NRe}(\varrho;\{\Pi_{b}\})\ge\frac{1}{2}\sup_{B\in\mathbb{H}(\mathcal{H}|\{\Pi_{b}\})}\sup_{A\in\mathbb{H}(\mathcal{H})}\big|{\rm Tr}\{[\tilde{A},\tilde{B}]_-\varrho\}\big|,
\label{uncertainty relation for the KD nonreality relative to a PVM proof step 2}
\end{eqnarray} 
where the supremum are now taken over the set $\mathbb{H}(\mathcal{H}|\{\Pi_{b}\})$ of all bounded Hermitian operators $B$ on $\mathcal{H}$ whose complete set of eigenprojectors is given by the PVM basis $\{\Pi_{b}\}$, and over the set $\mathbb{H}(\mathcal{H})$ of all bounded Hermitian operator $A$ on $\mathcal{H}$. Combining Eqs. (\ref{uncertainty relation for the KD nonreality relative to a PVM proof step 1}) and (\ref{uncertainty relation for the KD nonreality relative to a PVM proof step 2}), we thus finally obtain 
\begin{eqnarray}
&&\mathcal{Q}_{\rm KD}^{\rm NRe}(\varrho;\{\Pi_{a}\})\mathcal{Q}_{\rm KD}^{\rm NRe}(\varrho;\{\Pi_{b}\})\nonumber\\
&\ge&\frac{1}{4}\sup_{A\in\mathbb{H}(\mathcal{H}|\{\Pi_{a}\})}\sup_{B\in\mathbb{H}(\mathcal{H}\})}\big|{\rm Tr}\{[\tilde{A},\tilde{B}]_-\varrho\}\big|\nonumber\\
&\times&\sup_{B\in\mathbb{H}(\mathcal{H}|\{\Pi_{b}\})}\sup_{A\in\mathbb{H}(\mathcal{H})}\big|{\rm Tr}\{[\tilde{A},\tilde{B}]_-\varrho\}\big|\nonumber\\
&\ge&\frac{1}{4}\sup_{A\in\mathbb{H}(\mathcal{H}|\{\Pi_{a}\})}\sup_{B\in\mathbb{H}(\mathcal{H}|\{\Pi_{b}\})}\big|{\rm Tr}\{[\tilde{A},\tilde{B}]_-\varrho\}\big|^2,
\label{uncertainty relation for the KD nonreality relative to a PVM proof step 3}
\end{eqnarray} 
where to get the inequality in Eq. (\ref{uncertainty relation for the KD nonreality relative to a PVM proof step 3}), we have made use of the fact that $\sup_{X\in\mathbb{H}(\mathcal{H})}\{\cdot\}\ge\sup_{X\in\mathbb{H}(\mathcal{H}|\{\Pi_x\})}\{\cdot\}$. \qed

One can see that the lower bound in the trade-off relation of Eq. (\ref{uncertainty relation for the KD nonreality relative to a PVM}) takes a form similar to that of the Robertson uncertainty relation. Unlike the latter, however, it involves optimizations over the convex sets $\mathbb{H}(\mathcal{H}|\{\Pi_{a}\})$ and $\mathbb{H}(\mathcal{H}|\{\Pi_{b}\})$ of all Hermitian operators on $\mathcal{H}$ whose complete set of eigenprojectors are given respectively by the PVM bases $\{\Pi_a\}$ and $\{\Pi_b\}$ relative to which we define the KD nonreality in the state $\varrho$: $\mathcal{Q}_{\rm KD}^{\rm NRe}(\varrho;\{\Pi_{a}\})$ and $\mathcal{Q}_{\rm KD}^{\rm NRe}(\varrho;\{\Pi_{b}\})$. The trade-off relation shows that if there is a pair of Hermitian operators $A\in\mathbb{H}(\mathcal{H}|\{\Pi_{a}\})$ and $B\in\mathbb{H}(\mathcal{H}|\{\Pi_{b}\})$ such that ${\rm Tr}\{[\tilde{A},\tilde{B}]_-\varrho\}\neq 0$, then the lower bound in Eq. (\ref{uncertainty relation for the KD nonreality relative to a PVM}) is not vanishing. In this case, both the KD nonreality $\mathcal{Q}_{\rm KD}^{\rm NRe}(\varrho;\{\Pi_{a}\})$ in $\varrho$ relative to the PVM basis $\{\Pi_a\}$ and the KD nonreality $\mathcal{Q}_{\rm KD}^{\rm NRe}(\varrho;\{\Pi_{b}\})$ in $\varrho$ relative to the PVM basis $\{\Pi_b\}$ cannot be vanishing, and their product must satisfy the trade off relation of Eq. (\ref{uncertainty relation for the KD nonreality relative to a PVM}).  

Let us proceed to show that the lower bounds in Eqs. (\ref{KD nonreality relative to a PVM is lower bounded by a normalized Robertson bound}) and (\ref{KD nonreality is lower bounded by a normalized trace-norm asymmetry}) and the trade-off relation of Eq. (\ref{uncertainty relation for the KD nonreality relative to a PVM}) for the KD nonreality in a state relative to a rank-1 orthogonal PVM basis, imply lower bounds and trade-off relation for the $l_1$-coherence of the state relative to the orthonormal basis corresponding to the PVM basis. First, note that as shown in Ref. \cite{Agung KD-nonreality coherence}, the KD nonreality $\mathcal{Q}_{\rm KD}^{\rm NRe}(\varrho;\{\Pi_{a}\})$ in the state $\varrho$ relative to the rank-1 PVM basis $\{\Pi_a\}$ gives a lower bound to the $l_1$-norm coherence $C_{l_1}(\varrho;\{\ket{a}\})$ of $\varrho$ relative to the orthonormal basis $\{\ket{a}\}$ corresponding to $\{\Pi_a\}$ as
\begin{eqnarray}
C_{l_1}(\varrho;\{\ket{a}\})\ge\mathcal{Q}_{\rm KD}^{\rm NRe}(\varrho;\{\Pi_{a}\}). 
\label{l1-norm coherence is lower bounded by the KD nonreality relative to a PVM}
\end{eqnarray}
Moreover, for an arbitrary state of a single qubit and arbitrary orthonormal basis $\{\ket{a}\}$ of $\mathbb{C}^2$, the inequality becomes equality \cite{Agung KD-nonreality coherence}. 

Using Eqs. (\ref{KD nonreality relative to a PVM is lower bounded by a normalized Robertson bound}) and (\ref{l1-norm coherence is lower bounded by the KD nonreality relative to a PVM}), we thus obtain the following result. \\
{\bf Corollary 1}. The $l_1$-norm coherence of a quantum state $\varrho$ on a finite-dimensional Hilbert space $\mathcal{H}$ relative to an incoherent orthonormal basis $\{\ket{a}\}$ of $\mathcal{H}$ is lower bounded as:
\begin{eqnarray}
C_{l_1}(\varrho;\{\ket{a}\})&\ge&\frac{1}{2}\sup_{A\in\mathbb{H}(\mathcal{H}|\{\Pi_{a}\})}\sup_{B\in\mathbb{H}(\mathcal{H})}\big|{\rm Tr}\{\tilde{B}[\tilde{A},\varrho]_-\}\big|.
%&=&\sup_{A\in\mathbb{H}(\mathcal{H}|\{\Pi_{a}\})}\|[A,\varrho]_-\|_1/2\|A\|_{\infty}\nonumber\\
%&=&\sup_{A\in\mathbb{H}(\mathcal{H}|\{\Pi_{a}\})}\mathcal{A}_{\rm Tr}(A;\varrho)/\|A\|_{\infty}. 
\label{l1-norm coherence is lower bounded by a normalized Robertson bound}
\end{eqnarray}

As shown in Appendix \ref{Proof that the equality in lemma 1 is achieved for a single qubit}, for two-dimensional Hilbert space $\mathbb{C}^2$, the inequality in Eq. (\ref{l1-norm coherence is lower bounded by a normalized Robertson bound}) becomes equality for arbitrary single qubit state and arbitrary orthonormal basis, as expressed in Eq. (\ref{KD-nonreality for two-dimensional Hilbert space}). 

Next, from Eqs. (\ref{KD nonreality is lower bounded by a normalized trace-norm asymmetry}) and (\ref{l1-norm coherence is lower bounded by the KD nonreality relative to a PVM}), we have the following result.\\
{\bf Corollary 2}. The KD nonreality in a quantum state $\varrho$ on a finite-dimensional Hilbert space $\mathcal{H}$ relative to a rank-1 orthogonal PVM basis $\{\Pi_a\}$ of $\mathcal{H}$, the corresponding $l_1$-norm coherence of $\varrho$ relative to the orthonormal basis $\{\ket{a}\}$, and the trace-norm asymmetry of $\varrho$ relative to the translation group generated by any Hermitian operator $A$ with a complete set of eigenprojectors $\{\Pi_a\}$, obey the following ordering:
\begin{eqnarray}
C_{l_1}(\varrho;\{\ket{a}\})&\ge&\mathcal{Q}_{\rm KD}^{\rm NRe}(\varrho;\{\Pi_{a}\})\ge\sup_{A\in\mathbb{H}(\mathcal{H}|\{\Pi_{a}\})}\|[\tilde{A},\varrho]_-\|_1/2. 
\label{l1-norm coherence vs KD nonreality vs trace-norm asymmetry}
\end{eqnarray} 

For two-dimensional Hilbert space $\mathbb{C}^2$, as shown in Appendix \ref{Proof that the equality in lemma 1 is achieved for a single qubit}, both inequalities in Eq. (\ref{l1-norm coherence vs KD nonreality vs trace-norm asymmetry}) become equalities for arbitrary state $\varrho$ and arbitrary incoherent orthonormal basis $\{\ket{a}\}$.

Recall that whilst $C_{l_1}(\varrho;\{\ket{a}\})$ is a measure of quantum coherence of $\varrho$ which is independent of its encoding in the reference incoherent orthonormal basis $\{\ket{a}\}$, the trace-norm asymmetry $\|[\tilde{A},\varrho]\|_1/2$ can also be seen as a measure of coherence (as translational asymmetry) of $\varrho$ which depends on its encoding in the reference incoherent eigenbasis $\{\ket{a}\}$ of the generator $\tilde{A}=A/\|A\|_{\infty}$ of the translation group $U_{\theta}=e^{-i\tilde{A}\theta}$, $\theta\in\mathbb{R}$. The former is sometimes called speakable coherence while the latter is called unspeakable coherence \cite{Marvian - Spekkens speakable and unspeakable coherence}.   

Finally, combining Eqs. (\ref{uncertainty relation for the KD nonreality relative to a PVM}) with (\ref{l1-norm coherence is lower bounded by the KD nonreality relative to a PVM}), we obtain the following trade-off relation.\\
{\bf Corollary 3}. The $l_1$-norm coherence of a quantum state $\varrho$ on a finite-dimensional Hilbert space $\mathcal{H}$ relative to an incoherent orthonormal basis $\{\ket{a}\}$ of $\mathcal{H}$ and that relative to an incoherent orthonormal basis $\{\ket{b}\}$ of $\mathcal{H}$ satisfy the following trade-off relation:
\begin{eqnarray}
&&C_{l_1}(\varrho;\{\ket{a}\})C_{l_1}(\varrho;\{\ket{b}\})\ge\frac{1}{4}\sup_{A\in\mathbb{H}(\mathcal{H}|\{\Pi_{a}\})}\sup_{B\in\mathbb{H}(\mathcal{H}|\{\Pi_{b}\})}\big|{\rm Tr}\{[\tilde{A},\tilde{B}]_-\varrho\}\big|^2. 
\label{uncertainty relation for l1-norm coherence}
\end{eqnarray}  

Next, from Eq. (\ref{KD nonreality relative to a PVM is lower bounded by a normalized noncommutativity}) of Proposition 1, we obtain the following additive trade-off relation for the KD nonreality in a state $\varrho$ relative to the PVM basis $\{\Pi_a\}$ and that relative to the PVM basis $\{\Pi_b\}$:
\begin{eqnarray}
&&\mathcal{Q}_{\rm KD}^{\rm NRe}(\varrho;\{\Pi_{a}\})+\mathcal{Q}_{\rm KD}^{\rm NRe}(\varrho;\{\Pi_{b}\})\ge\sup_{A\in\mathbb{H}(\mathcal{H}|\{\Pi_{a}\})}\sup_{B\in\mathbb{H}(\mathcal{H}|\{\Pi_{b}\})}\big|{\rm Tr}\{[\tilde{A},\tilde{B}]_-\varrho\}\big|. 
\label{additive uncertainty relation for KD-nonreality coherence}
\end{eqnarray}
The proof follows exactly similar steps as the proof of Eq. (\ref{uncertainty relation for the KD nonreality relative to a PVM}) of Proposition 3. It can also be proven by applying the inequality for the arithmetic mean and geometric mean, i.e., $(a+b)/2\ge \sqrt{ab}$, $a,b\in\mathbb{R}^+$, to Eq. (\ref{uncertainty relation for the KD nonreality relative to a PVM}). Since $\mathcal{Q}_{\rm KD}^{\rm NRe}(\varrho;\{\Pi_{a}\})$ is a faithful measure of coherence, Eq. (\ref{additive uncertainty relation for KD-nonreality coherence}) has a form of additive uncertainty relation for coherence measure reported in Refs. \cite{Korzekwa quantum-classical decomposition,Singh uncertainty relation for coherence,Yuan uncertainty relation for coherence,Hall quantum-classical decomposition}. One can then check that the left-hand side is not vanishing when the state is not totally mixed, i.e., $\varrho\neq\mathbb{I}/d$, and the PVM bases are noncommuting as stated in Theorem 1 of Ref. \cite{Yuan uncertainty relation for coherence}. In particular, combining Eq. (\ref{additive uncertainty relation for KD-nonreality coherence}) with Eq. (\ref{l1-norm coherence is lower bounded by the KD nonreality relative to a PVM}), we have 
\begin{eqnarray}
&&C_{l_1}(\varrho;\{\ket{a}\})+C_{l_1}(\varrho;\{\ket{b}\})\ge\sup_{A\in\mathbb{H}(\mathcal{H}|\{\Pi_{a}\})}\sup_{B\in\mathbb{H}(\mathcal{H}|\{\Pi_{b}\})}\big|{\rm Tr}\{[\tilde{A},\tilde{B}]_-\varrho\}\big|. 
\label{additive uncertainty relation for l1-norm coherence}
\end{eqnarray} 

We note that unlike the standard entropic uncertainty relation \cite{Coles entropic uncertainty relation review,Wehner entropic UR review}, the lower bound in Eqs. (\ref{additive uncertainty relation for KD-nonreality coherence}) and (\ref{additive uncertainty relation for l1-norm coherence}) depends on the state as for the uncertainty relation for coherence measures in Refs. \cite{Korzekwa quantum-classical decomposition,Singh uncertainty relation for coherence,Yuan uncertainty relation for coherence,Hall quantum-classical decomposition}. In particular, it is vanishing when the state is maximally mixed $\varrho=\mathbb{I}/d$, in case of which, the left-hand sides in Eqs. (\ref{additive uncertainty relation for KD-nonreality coherence}) and (\ref{additive uncertainty relation for l1-norm coherence}) are also vanishing. Hence, the uncertainty relation depends on the purity of the state as expected \cite{Korzekwa quantum-classical decomposition}. It is interesting in the future to compare the type of lower bound in Eqs. (\ref{additive uncertainty relation for KD-nonreality coherence}) and (\ref{additive uncertainty relation for l1-norm coherence}) to those reported in Refs. \cite{Korzekwa quantum-classical decomposition,Singh uncertainty relation for coherence,Yuan uncertainty relation for coherence,Hall quantum-classical decomposition}. In the Appendix \ref{Additive l1 norm trade-off relation for a single qubit}, we evaluate the optimization in the lower bound analytically for two-dimensional system, showing that it is determined by the purity of the state, and three parameters that characterize the pairwise noncommutativity among the two PVM bases and the eigenbasis of the state. We furthermore show that for pure state in two dimensional Hilbert space, the inequality becomes equality when the bases $\{\ket{a}\}$, $\{\ket{b}\}$, and the eigenbasis of $\varrho$ comprise a set of three mutually unbiased bases of $\mathbb{C}^2$.  

\subsection{Lower bound and uncertainty relation for the KD nonclassicality in a state relative to a PVM}

First, using Lemma 2, we obtain the following proposition. \\
{\bf Proposition 4}. The KD nonclassicality in a state $\varrho$ on a finite-dimensional Hilbert space $\mathcal{H}$ relative to a PVM basis $\{\Pi_{a}\}$ of $\mathcal{H}$ defined in Eq. (\ref{KD nonclassicality relative to a PVM}) is lower bounded as
\begin{eqnarray}
\mathcal{Q}_{\rm KD}^{\rm NCl}(\varrho;\{\Pi_{a}\})
&\ge&\frac{1}{2}\sup_{A\in\mathbb{H}(\mathcal{H}|\{\Pi_{a}\})}\sup_{B\in\mathbb{H}(\mathcal{H})}\big\{\big(\big|{\rm Tr}\{\varrho[\tilde{A}_{\varrho},\tilde{B}_{\varrho}]_-\}\big|^2\nonumber\\
&+&\big|{\rm Tr}\{\varrho[\tilde{A}_{\varrho},\tilde{B}_{\varrho}]_+\}-2{\rm Tr}\{\tilde{A}_{\varrho}\varrho\}{\rm Tr}\{\tilde{B}_{\varrho}\varrho\}\big|^2\big)^{1/2}\big\}-1. 
\label{KD-nonclassility relative to a PVM is lower bounded by Robertson-Schrodinger bound 1}
\end{eqnarray}
{\bf Proof}. Taking the supremum over the set $\mathcal{M}_{\rm r1PVM}(\mathcal{H})$ of all the rank-1 PVM bases $\{\Pi_{b}\}$ of $\mathcal{H}$ to both sides of Eq. (\ref{KD nonclassility relative to a pair of POVMs is lower bounded by Robertson-Schrodinger bound}), and noting Eq. (\ref{KD nonclassicality relative to a PVM}), we obtain
\begin{eqnarray}
\mathcal{Q}_{\rm KD}^{\rm NCl}(\varrho;\{\Pi_{a}\})
&=&\sup_{\{\Pi_{b}\}\in\mathcal{M}_{\rm r1PVM}(\mathcal{H})}{\rm NCl}(\{{\rm Pr}_{\rm KD}(a,b|\varrho)\})\nonumber\\
&\ge&\frac{1}{2}\sup_{B\in\mathbb{H}(\mathcal{H}|\{b\})}\big\{\big(\big|{\rm Tr}\{\varrho[\tilde{A}_{\varrho},\tilde{B}_{\varrho}]_-\}\big|^2\nonumber\\
&+&\big|{\rm Tr}\{\varrho[\tilde{A}_{\varrho},\tilde{B}_{\varrho}]_+\}-2{\rm Tr}\{\tilde{A}_{\varrho}\varrho\}{\rm Tr}\{\tilde{B}_{\varrho}\varrho\}\big|^2\big)^{1/2}\big\}-1.   
\label{KD nonclassicality relative to a PVM is lower bounded by Robertson bound 1 step 4}
\end{eqnarray}
Observe further that the left-hand side depends only on the PVM basis $\{\Pi_a\}$, i.e., it is independent of the eigenvalues spectrum $\{a\}$ of $A$ and the eigenvalues spectrum $\{b\}$ of $B$. Noting this, the inequality of Eq. (\ref{KD nonclassicality relative to a PVM is lower bounded by Robertson bound 1 step 4}) can be further strengthened to get Eq. (\ref{KD-nonclassility relative to a PVM is lower bounded by Robertson-Schrodinger bound 1}). \qed

We then obtain the following trade-off relation. \\
{\bf Proposition 5}. The KD nonclassicality in a quantum state $\varrho$ on a finite-dimensional Hilbert space $\mathcal{H}$ relative to a PVM basis $\{\Pi_a\}$ of $\mathcal{H}$ and that relative to a PVM basis $\{\Pi_b\}$ of $\mathcal{H}$ satisfy the following trade-off relation:
\begin{eqnarray}
&&(\mathcal{Q}_{\rm KD}^{\rm NCl}(\varrho;\{\Pi_{a}\})+1)(\mathcal{Q}_{\rm KD}^{\rm NCl}(\varrho;\{\Pi_{b}\})+1)\nonumber\\
&\ge&\frac{1}{4}\sup_{A\in\mathbb{H}(\mathcal{H}|\{\Pi_{a}\})}\sup_{B\in\mathbb{H}(\mathcal{H}|\{\Pi_{b}\})}\big\{\big|{\rm Tr}\{\varrho[\tilde{A}_{\varrho},\tilde{B}_{\varrho}]_-\}\big|^2\nonumber\\
&+&\big|{\rm Tr}\{\varrho[\tilde{A}_{\varrho},\tilde{B}_{\varrho}]_+\}-2{\rm Tr}\{\tilde{A}_{\varrho}\varrho\}{\rm Tr}\{\tilde{B}_{\varrho}\varrho\}\big|^2\big\}.
\label{uncertainty relation for the KD classicality relative to a PVM}
\end{eqnarray} \\
{\bf Proof}. First, swapping the role of $A$ and $B$ in Eq. (\ref{KD-nonclassility relative to a PVM is lower bounded by Robertson-Schrodinger bound 1}), we have 
\begin{eqnarray}
&&\mathcal{Q}_{\rm KD}^{\rm NCl}(\varrho;\{\Pi_{b}\})+1\nonumber\\
&\ge&\frac{1}{2}\sup_{B\in\mathbb{H}(\mathcal{H}|\{\Pi_{b}\})}\sup_{A\in\mathbb{H}(\mathcal{H})}\big\{\big(|{\rm Tr}\{\varrho[\tilde{A}_{\varrho},\tilde{B}_{\varrho}]_-\}|^2\nonumber\\
&+&\big|{\rm Tr}\{\varrho[\tilde{A}_{\varrho},\tilde{B}_{\varrho}]_+\}-2{\rm Tr}\{\tilde{A}_{\varrho}\varrho\}{\rm Tr}\{\tilde{B}_{\varrho}\varrho\}\big|^2\big)^{1/2}\big\}. 
\label{KD-nonclassility relative to a PVM is lower bounded by Robertson-Schrodinger bound 2}
\end{eqnarray} 
Hence, combining Eqs. (\ref{KD-nonclassility relative to a PVM is lower bounded by Robertson-Schrodinger bound 1}) and (\ref{KD-nonclassility relative to a PVM is lower bounded by Robertson-Schrodinger bound 2}), we obtain 
\begin{eqnarray}
&&(\mathcal{Q}_{\rm KD}^{\rm NCl}(\varrho;\{\Pi_{a}\})+1)(\mathcal{Q}_{\rm KD}^{\rm NCl}(\varrho;\{\Pi_{b}\})+1)\nonumber\\
&\ge&\frac{1}{4}\Big(\sup_{A\in\mathbb{H}(\mathcal{H}|\{\Pi_{a}\})}\sup_{B\in\mathbb{H}(\mathcal{H})}\big\{\big(|{\rm Tr}\{\varrho[\tilde{A}_{\varrho},\tilde{B}_{\varrho}]_-\}|^2\nonumber\\
&+&\big|{\rm Tr}\{\varrho[\tilde{A}_{\varrho},\tilde{B}_{\varrho}]_+\}-2{\rm Tr}\{\tilde{A}_{\varrho}\varrho\}{\rm Tr}\{\tilde{B}_{{\varrho}}\varrho\}\big|^2\big)^{1/2}\big\}\Big)\nonumber\\
&\times&\Big(\sup_{B\in\mathbb{H}(\mathcal{H}|\{\Pi_{b}\})}\sup_{A\in\mathbb{H}(\mathcal{H})}\big\{\big(|{\rm Tr}\{\varrho[\tilde{A}_{\varrho},\tilde{B}_{\varrho}]_-\}|^2\nonumber\\
&+&\big|{\rm Tr}\{\varrho[\tilde{A}_{\varrho},\tilde{B}_{\varrho}]_+\}-2{\rm Tr}\{\tilde{A}_{\varrho}\varrho\}{\rm Tr}\{\tilde{B}_{\varrho}\varrho\}\big|^2\big)^{1/2}\big\}\Big)\nonumber\\
&\ge&\frac{1}{4}\sup_{A\in\mathbb{H}(\mathcal{H}|\{\Pi_{a}\})}\sup_{B\in\mathbb{H}(\mathcal{H}|\{\Pi_{b}\})}\big\{|{\rm Tr}\{\varrho[\tilde{A}_{\varrho},\tilde{B}_{\varrho}]_-\}|^2\nonumber\\
&+&\big|{\rm Tr}\{\varrho[\tilde{A}_{\varrho},\tilde{B}_{\varrho}]_+\}-2{\rm Tr}\{\tilde{A}_{\varrho}\varrho\}{\rm Tr}\{\tilde{B}_{\varrho}\varrho\}\big|^2\big\},
\label{uncertainty relation for the KD nonclassicality relative to a PVM proof step 1}
\end{eqnarray}  
where the last inequality in Eq. (\ref{uncertainty relation for the KD nonclassicality relative to a PVM proof step 1}) is due to the fact that $\sup_{X\in\mathbb{H}(\mathcal{H})}\{\cdot\}\ge\sup_{X\in\mathbb{H}(\mathcal{H}|\{\Pi_{x}\})}\{\cdot\}$.  \qed

One can see that Eq. (\ref{uncertainty relation for the KD classicality relative to a PVM}) takes a form analogous to the Robertson-Schr\"odinger uncertainty relation for observables $\tilde{A}_{\varrho}$ and $\tilde{B}_{\varrho}$. Unlike the Robertson-Schr\"odinger uncertainty relation, however, the lower bound in Eq. (\ref{uncertainty relation for the KD classicality relative to a PVM}) is nonlinear in the state $\varrho$. Moreover, there are optimizations over a pair of convex sets $\mathbb{H}(\mathcal{H}|\{\Pi_a\})$ and $\mathbb{H}(\mathcal{H}|\{\Pi_b\})$ of Hermitian operators whose complete sets of eigenprojectors are respectively $\{\Pi_a\}$ and $\{\Pi_b\}$ relative to which we define the KD nonclassicality in $\varrho$: $\mathcal{Q}_{\rm KD}^{\rm NCl}(\varrho;\{\Pi_{a}\})$ and $\mathcal{Q}_{\rm KD}^{\rm NCl}(\varrho;\{\Pi_{b}\})$. Recall that the KD nonclassicality in a state relative to a PVM basis quantifies the total quantumness, i.e., it quantifies simultaneously the nonreality and negativity of the corresponding KD quasiprobability, capturing the noncommutativity between the state and the PVM basis. In this sense, the trade-off relation of Eq. (\ref{uncertainty relation for the KD classicality relative to a PVM}) thus imposes a restriction on a joint nonclassicality in a quantum state relative to a pair of noncommuting PVM bases. 

Let us proceed to show that the lower bound and the trade-off relation for the KD nonclassicality relative to a PVM basis obtained above lead also to a lower bound and a trade-off relation for the corresponding $l_1$-norm coherence. First, as shown in Appendix \ref{Proof that KD nonclassicality coherence gives a lower bound to the l1-norm coherence}, the KD nonclassicality in a state $\varrho$ on a finite-dimensional Hilbert space $\mathcal{H}$ relative to a rank-1 orthogonal PVM basis $\{\Pi_a\}$ of $\mathcal{H}$ gives a lower bound to the $l_1$-norm coherence of the state $\varrho$ relative to the incoherent orthonormal basis $\{\ket{a}\}$ corresponding to $\{\Pi_a\}$, i.e., 
\begin{eqnarray}
C_{l_1}(\varrho;\{\ket{a}\})\ge\mathcal{Q}_{\rm KD}^{\rm NCl}(\varrho;\{\Pi_{a}\}). 
\label{l1-norm coherence is lower bounded by the KD nonclassicality relative to a PVM}
\end{eqnarray}

Combining Eq. (\ref{l1-norm coherence is lower bounded by the KD nonclassicality relative to a PVM})  with Eq. (\ref{KD-nonclassility relative to a PVM is lower bounded by Robertson-Schrodinger bound 1}), we thus obtain the following corollary. \\
{\bf Corollary 4}. The $l_1$-norm coherence of a quantum state $\varrho$ on a finite-dimensional Hilbert space $\mathcal{H}$ relative to an incoherent orthonormal basis $\{\ket{a}\}$ of $\mathcal{H}$ is lower bounded as
\begin{eqnarray}
C_{l_1}(\varrho;\{\ket{a}\})&\ge&\frac{1}{2}\sup_{A\in\mathbb{H}(\mathcal{H}|\{\Pi_{a}\})}\sup_{B\in\mathbb{H}(\mathcal{H})}\big\{\big(|{\rm Tr}\{\varrho[\tilde{A}_{\varrho},\tilde{B}_{\varrho}]_-\}|^2\nonumber\\
&+&\big|{\rm Tr}\{\varrho[\tilde{A}_{\varrho},\tilde{B}_{\varrho}]_+\}-2{\rm Tr}\{\tilde{A}_{\varrho}\varrho\}{\rm Tr}\{\tilde{B}_{\varrho}\varrho\}\big|^2\big)^{1/2}\big\}-1.
\label{l1-norm coherence is lower bounded by reqularized Robertson-Schrodinger bound}
\end{eqnarray} 

Next, combining Eq. (\ref{l1-norm coherence is lower bounded by the KD nonclassicality relative to a PVM}) with Eq. (\ref{uncertainty relation for the KD classicality relative to a PVM}), we obtain the following trade-off relation. \\
{\bf Corollary 5}. The $l_1$-norm coherence of a quantum state $\varrho$ on a finite-dimensional Hilbert space $\mathcal{H}$ relative to an orthonormal basis $\{\ket{a}\}$ of $\mathcal{H}$ and that relative to an orthonormal basis $\{\ket{b}\}$ of $\mathcal{H}$ satisfy the following trade-off relation:
\begin{eqnarray}
&&\big(C_{l_1}(\varrho;\{\ket{a}\})+1\big)\big(C_{l_1}(\varrho;\{\ket{b}\})+1\big)\nonumber\\
&\ge&\frac{1}{4}\sup_{A\in\mathbb{H}(\mathcal{H}|\{\Pi_{a}\})}\sup_{B\in\mathbb{H}(\mathcal{H}|\{\Pi_{b}\})}\big\{\big(|{\rm Tr}\{\varrho[\tilde{A}_{\varrho},\tilde{B}_{\varrho}]_-\}|^2\nonumber\\
&+&\big|{\rm Tr}\{\varrho[\tilde{A}_{\varrho},\tilde{B}_{\varrho}]_+\}-2{\rm Tr}\{\tilde{A}_{\varrho}\varrho\}{\rm Tr}\{\tilde{B}_{\varrho}\varrho\}\big|^2\big)\big\}.
\label{Robertson-Schrodinger uncertainty relation for l1-norm coherence}
\end{eqnarray} 

Following exactly similar steps as above, we can also prove the following additive trade-off relation for the $l_1$-norm coherence of a state $\varrho$ relative to an orthonormal basis $\{\ket{a}\}$ and that relative to an orthonormal basis $\{\ket{b}\}$ as:
\begin{eqnarray}
&&C_{l_1}(\varrho;\{\ket{a}\})+C_{l_1}(\varrho;\{\ket{b}\})\nonumber\\
&\ge&\sup_{A\in\mathbb{H}(\mathcal{H}|\{\Pi_{a}\})}\sup_{B\in\mathbb{H}(\mathcal{H}|\{\Pi_{b}\})}\big\{\big(|{\rm Tr}\{\varrho[\tilde{A}_{\varrho},\tilde{B}_{\varrho}]_-\}|^2\nonumber\\
&+&\big|{\rm Tr}\{\varrho[\tilde{A}_{\varrho},\tilde{B}_{\varrho}]_+\}-2{\rm Tr}\{\tilde{A}_{\varrho}\varrho\}{\rm Tr}\{\tilde{B}_{\varrho}\varrho\}\big|^2\big)^{1/2}\big\}-2.
\label{Robertson-Schrodinger uncertainty relation for l1-norm coherence}
\end{eqnarray} 
%We again leave for future study the detailed comparison between the lower bound in Eq. (\ref{Robertson-Schrodinger uncertainty relation for l1-norm coherence}) and those reported in Refs. \cite{Korzekwa quantum-classical decomposition,Singh uncertainty relation for coherence,Yuan uncertainty relation for coherence,Hall quantum-classical decomposition}.

\section{Operational and statistical meaning}

In this section we discuss operational and information theoretical interpretations of the KD nonreality and KD nonclassicality in a state relative to a PVM basis in terms of transparent laboratory operations. One is based on the representation of the KD quasiprobability in terms of weak value which can be obtained using various methods in experiment, and the other is based on the decomposition of the KD quasiprobability in terms of real and nonnegative joint probability and quantum modification terms obtained via two successive projective measurements. 

First, one observes that the KD nonreality and the KD nonclassicality in a state $\varrho$ relative to a PVM basis $\{\Pi_a\}$ defined respectively in Eqs. (\ref{KD nonreality relative to a PVM}) and (\ref{KD nonclassicality relative to a PVM}) can be expressed as 
\begin{eqnarray}
\label{KD nonreality relative to a PVM as average absolute imaginary part of weak value}
\mathcal{Q}_{\rm KD}^{\rm NRe}(\varrho;\{\Pi_{a}\})&=&\sup_{\{\Pi_{b}\}\in\mathcal{M}_{\rm r1PVM}(\mathcal{H})}
\sum_{a,b}\big|{\rm Im}\pi_a^{\rm w}(\Pi_{b}|\varrho)\big|{\rm Tr}\{\Pi_b\varrho\},\\
\label{KD nonclassicality relative to a PVM as average absolute weak value}
\mathcal{Q}_{\rm KD}^{\rm NCl}(\varrho;\{\Pi_{a}\})&=&\sup_{\{\Pi_{b}\}\in\mathcal{M}_{\rm r1PVM}(\mathcal{H})}\sum_{a,b}\big|\pi_a^{\rm w}(\Pi_{b}|\varrho)\big|{\rm Tr}\{\Pi_b\varrho\}-1. 
\end{eqnarray}
Here, $\pi_a^{\rm w}(\Pi_{b}|\varrho):=\frac{{\rm Tr}\{\Pi_b\Pi_a\varrho\}}{{\rm Tr}\{\Pi_b\varrho\}}$ is known as the weak value of $\Pi_a$ with the preselected state $\varrho$ and postselected state $\ket{b}$ \cite{Aharonov weak value,Aharonov-Daniel book,Wiseman weak value}. It is in general complex and its real part may lie outside $[0,1]$. Remarkably, the real and imaginary parts of the weak value can be estimated in experiment without recourse to state tomography either using weak measurement with postselection \cite{Aharonov weak value,Aharonov-Daniel book,Wiseman weak value,Lundeen complex weak value,Jozsa complex weak value,Lundeen measurement of KD distribution,Salvail direct measurement KD distribution,Bamber measurement of KD distribution,Thekkadath measurement of density matrix} or different methods without weak measurement \cite{Johansen quantum state from successive projective measurement,Vallone strong measurement to reconstruct quantum wave function,Cohen estimating of weak value with strong measurements,Lostaglio KD quasiprobability and quantum fluctuation,Wagner measuring weak values and KD quasiprobability,Hernandez-Gomez experimental observation of TBMH negativity}. Noting this, the KD nonreality and KD nonclassicality in a state relative to a PVM basis of Eqs. (\ref{KD nonreality relative to a PVM as average absolute imaginary part of weak value}) and (\ref{KD nonclassicality relative to a PVM as average absolute weak value}) can thus be directly operationally estimated using weak value measurement together with the classical optimization over the set $\mathcal{M}_{\rm r1PVM}(\mathcal{H})$ of all the rank-1 orthogonal PVM bases of the Hilbert space $\mathcal{H}$. This estimation scheme should in principle be implementable in terms of variational quantum circuits using the currently available NISQ hardware \cite{Cerezo VQA review}. 

The above operational interpretation suggests the following information theoretical meaning of the KD nonreality $\mathcal{Q}_{\rm KD}^{\rm NRe}(\varrho;\{\Pi_{a}\})$ in a state $\varrho$ relative to a PVM basis $\{\Pi_{a}\}$ defined in Eq. (\ref{KD nonreality relative to a PVM}) and the associated trade-off relation expressed in Eq. (\ref{uncertainty relation for the KD nonreality relative to a PVM}). First, applying the Jensen inequality to Eq. (\ref{KD nonreality relative to a PVM as average absolute imaginary part of weak value}), we have 
\begin{eqnarray}
\mathcal{Q}_{\rm KD}^{\rm NRe}(\varrho;\{\Pi_{a}\})^2&=&\big(\sup_{\{\Pi_{b}\}\in\mathcal{M}_{\rm r1PVM}(\mathcal{H})}
\sum_{a,b}\big|{\rm Im}\pi_a^{\rm w}(\Pi_{b}|\varrho)\big|{\rm Tr}\{\Pi_b\varrho\}\big)^2\nonumber\\
&\le&\sup_{\{\Pi_{b}\}\in\mathcal{M}_{\rm r1PVM}(\mathcal{H})}\sum_{a,b}\big|{\rm Im}\pi_a^{\rm w}(\Pi_{b}|\varrho)\big|^2{\rm Tr}\{\Pi_b\varrho\}\nonumber\\
&:=&\epsilon^2_{\{\Pi_a\}}(\varrho),
\label{KD nonreality relative to a PVM is equal to the total standard deviation of the estimation of PVM in the worst case scenario}
\end{eqnarray}
where $\epsilon^2_{\{\Pi_a\}}(\varrho)$ is the total sum of the variance of the imaginary part of the weak value $\pi_a^{\rm w}(\Pi_{b}|\varrho)$ over the probability ${\rm Pr}(b|\varrho)={\rm Tr}\{\Pi_b\varrho\}$ maximized over the set $\mathcal{M}_{\rm r1PVM}(\mathcal{H})$ of all the PVM bases $\{\Pi_b\}$ of $\mathcal{H}$. On the other hand, it was argued by Johansen and Hall in Refs. \cite{Johansen weak value best estimation,Hall prior information}, that the variance of the imaginary part of the weak value $\pi^{\rm w}_a(\Pi_b|\varrho)$ over the probability ${\rm Pr}(b|\varrho)={\rm Tr}\{\Pi_b\varrho\}$ can be interpreted as the mean-squared error of the optimal estimation of $\Pi_a$ based on the outcomes of measurement described by the PVM basis $\{\Pi_b\}$ when the preparation is represented by the state $\varrho$. Noting this, $\epsilon^2_{\{\Pi_a\}}(\varrho)$ defined in Eq. (\ref{KD nonreality relative to a PVM is equal to the total standard deviation of the estimation of PVM in the worst case scenario}) may thus be statistically interpreted as the total mean-squared error of the optimal estimation of the PVM basis $\{\Pi_a\}$ based on projective measurement, given the preparation $\varrho$, in the worst case scenario. Equation (\ref{KD nonreality relative to a PVM is equal to the total standard deviation of the estimation of PVM in the worst case scenario}) thus shows that the total root-mean-squared error of the optimal estimation of the PVM basis $\{\Pi_a\}$ given $\varrho$ in the worst case scenario is lower bounded by the corresponding KD nonreality in $\varrho$ relative to the PVM basis $\{\Pi_a\}$. 

Combining Eq. (\ref{KD nonreality relative to a PVM is equal to the total standard deviation of the estimation of PVM in the worst case scenario}) with Eqs. (\ref{KD nonreality relative to a PVM is lower bounded by a normalized Robertson bound}) and (\ref{KD nonreality is lower bounded by a normalized trace-norm asymmetry}), we thus obtain the following results. \\
{\bf Corollary 5a}. The total root-mean-squared error of the optimal estimation of a PVM basis $\{\Pi_a\}$ of a finite-dimensional Hilbert space $\mathcal{H}$ given a preselected state $\varrho$ on $\mathcal{H}$, based on projective measurement described by a PVM basis in $\mathcal{M}_{\rm r1PVM}(\mathcal{H})$, in the worst case scenario, is lower bounded as 
\begin{eqnarray}
\epsilon_{\{\Pi_a\}}(\varrho)&\ge&\frac{1}{2}\sup_{A\in\mathbb{H}(\mathcal{H}|\{\Pi_{a}\})}\sup_{B\in\mathbb{H}(\mathcal{H})}\big|{\rm Tr}\{\tilde{B}[\tilde{A},\varrho]_-\}\big|.
%&=&\sup_{A\in\mathbb{H}(\mathcal{H}|\{\Pi_{a}\})}\mathcal{A}_{\rm Tr}(\tilde{A};\varrho).   
\label{lower bound for the MSEE of a PVM basis 1}
\end{eqnarray}
Moreover, it can also be lower bounded in terms of trace-norm asymmetry as 
\begin{eqnarray}
\epsilon_{\{\Pi_a\}}(\varrho)&\ge&\sup_{A\in\mathbb{H}(\mathcal{H}|\{\Pi_{a}\})}\|[A,\varrho]_-\|_1/2\|A\|_{\infty}.    
\label{lower bound for the MSEE of a PVM basis 2}
\end{eqnarray}

Next, combining Eqs. (\ref{KD nonreality relative to a PVM is equal to the total standard deviation of the estimation of PVM in the worst case scenario}) with (\ref{uncertainty relation for the KD nonreality relative to a PVM}), we have the following uncertainty relation. \\
{\bf Corollary 5b}. Given a preparation represented by a density operator $\varrho$ on a finite-dimensional Hilbert space $\mathcal{H}$, the total root-mean-squared errors of the optimal estimation of $\{\Pi_a\}$ of $\mathcal{H}$ based on projective measurement described by a PVM basis in $\mathcal{M}_{\rm r1PVM}(\mathcal{H})$, and that of the optimal estimation of $\{\Pi_b\}$ of $\mathcal{H}$, in the worst case scenario, satisfy the following trade-off relation:
\begin{eqnarray}
\epsilon_{\{\Pi_a\}}(\varrho)\epsilon_{\{\Pi_b\}}(\varrho)\ge\frac{1}{4}\sup_{A\in\mathbb{H}(\mathcal{H}|\{\Pi_{a}\})}\sup_{B\in\mathbb{H}(\mathcal{H}|\{\Pi_{b}\})}\big|{\rm Tr}\{[\tilde{A},\tilde{B}]_-\varrho\}\big|^2. 
\label{uncertainty relation for the total standard deviation of optimal estimation in the worst case scenario}
\end{eqnarray}

Let us proceed to discuss an operational interpretation of the KD nonclassicality in a state relative to a rank-1 PVM basis in terms of a sequence of two strong projective measurements. First, it has been shown by Johansen in Ref. \cite{Johansen quantum state from successive projective measurement} that the KD quasiprobability associated with a state $\varrho$ over a pair of rank-1 PVM bases $\{\Pi_a\}$ and $\{\Pi_b\}$ can be expressed as
\begin{eqnarray}
{\rm Pr}_{\rm KD}(a,b|\varrho)&=&{\rm Tr}\{\Pi_b\Pi_a\varrho\Pi_a\}+\frac{1}{2}{\rm Tr}\{(\varrho-\varrho_{\Pi_a})\Pi_b\}\nonumber\\
&-&i\frac{1}{2}{\rm Tr}\{(\varrho-\varrho_{\Pi_a})\Pi_{b|a}^{\pi/2}\}. 
\label{Johansen decomposition of KD quasiprobability}
\end{eqnarray}
Here, $\varrho_{\Pi_a}:=\Pi_a\varrho\Pi_a+(\mathbb{I}-\Pi_a)\varrho(\mathbb{I}-\Pi_a)$ is the state after a nonselective binary projective measurement described by $\{\Pi_a,\mathbb{I}-\Pi_a\}$, and $\Pi_{b|a}^{\pi/2}=e^{i\Pi_a\pi/2}\Pi_b e^{-i\Pi_a\pi/2}$. The first term on the right-hand side of Eq. (\ref{Johansen decomposition of KD quasiprobability}), i.e., ${\rm Tr}\{\Pi_b\Pi_a\varrho\Pi_a\}={\rm Tr}\{\Pi_b\frac{\Pi_a\varrho\Pi_a}{{\rm Tr}\{\varrho\Pi_a\}}\}{\rm Tr}\{\varrho\Pi_a\}$, is just the joint probability to get $a$ in the measurement described by $\{\Pi_a\}$ and then to get $b$ afterward in the measurement described by $\{\Pi_b\}$, so that it is always real and nonnegative. In this sense, the other two terms are called the quantum modification terms responsible for the negativity and nonreality of the KD-quasiprobability. One can then see that the negativity and the nonreality capture different forms of state disturbance due to the nonselective binary projective measurement $\{\Pi_a,\mathbb{I}-\Pi_a\}$ as captured by the expectation values of $\Pi_b$ and $\Pi_{b|a}^{\pi/2}$, respectively. 

Using the decomposition of the KD quasiprobability in Eq. (\ref{Johansen decomposition of KD quasiprobability}), the KD nonclassicality in $\varrho$ relative to the PVM $\{\Pi_a\}$ defined in Eq. (\ref{KD nonclassicality relative to a PVM}) can then be upper bounded as 
\begin{eqnarray}
\label{KD nonclassicality based on Johansen decomposition}
\mathcal{Q}_{\rm KD}^{\rm NCl}(\varrho;\{\Pi_{a}\})&=&\sup_{\{\Pi_{b}\}\in\mathcal{M}_{\rm r1PVM}(\mathcal{H})}\sum_{a,b}\big|{\rm Tr}\{\Pi_b\Pi_a\varrho\Pi_a\}+\frac{1}{2}{\rm Tr}\{(\varrho-\varrho_{\Pi_a})\Pi_b\}\nonumber\\
&-&i\frac{1}{2}{\rm Tr}\{(\varrho-\varrho_{\Pi_a})\Pi_{b|a}^{\pi/2}\}\big|-1\nonumber\\
&\le &\sup_{\{\Pi_{b}\}\in\mathcal{M}_{\rm r1PVM}(\mathcal{H})}\sum_{a,b}\big|{\rm Tr}\{(\varrho-\varrho_{\Pi_a})\Pi_b\}\big|:=\delta_{\{\Pi_a\}}(\varrho). 
%\label{state disturbance due binary projective measurement}
\end{eqnarray}
Here, to get Eq. (\ref{KD nonclassicality based on Johansen decomposition}), we have used triangle inequality, the normalization $\sum_{a,b}|{\rm Tr}\{\Pi_b\Pi_a\varrho\Pi_a\}|=\sum_{a,b}{\rm Tr}\{\Pi_b\Pi_a\varrho\Pi_a\}=1$ for any $\{\Pi_b\}\in\mathcal{M}_{\rm r1PVM}(\mathcal{H})$, and also $\sup_{\{\Pi_{b}\}\in\mathcal{M}_{\rm r1PVM}(\mathcal{H})}\sum_{a,b}|{\rm Tr}\{(\varrho-\varrho_{\Pi_a})\Pi_{b|a}^{\pi/2}\}|=\sup_{\{\Pi_{b}\}\in\mathcal{M}_{\rm r1PVM}(\mathcal{H})}\sum_{a,b}|{\rm Tr}\{(\varrho-\varrho_{\Pi_a})\Pi_b\}|$, where the equality is due to the fact that $\{\Pi_{b|a}^{\pi/2}\}$ comprises again a rank-1 PVM basis of $\mathcal{H}$, and the set of the PVM basis $\{\Pi_{b|a}^{\pi/2}\}$ is the same as the set of the PVM basis $\{\Pi_b\}$ given by $\mathcal{M}_{\rm r1PVM}(\mathcal{H})$. Hence, the KD nonclassicality $\mathcal{Q}_{\rm KD}^{\rm NCl}(\varrho;\{\Pi_{a}\})$ in $\varrho$ relative to the rank-1 PVM basis $\{\Pi_a\}$ gives a lower bound to the total disturbance $\delta_{\{\Pi_a\}}(\varrho)$ in the state $\varrho$ caused by the nonselective projective binary measurement $\{\Pi_a,\mathbb{I}-\Pi_a\}$ associated with the PVM basis $\{\Pi_a\}$. 

Combining Eq. (\ref{KD nonclassicality based on Johansen decomposition}) and Eq. (\ref{KD-nonclassility relative to a PVM is lower bounded by Robertson-Schrodinger bound 1}), we first have the following corollary. \\
{\bf Corollary 6a}. The total disturbance $\delta_{\{\Pi_a\}}(\varrho)$ in the state $\varrho$ caused by the nonselective projective binary measurement $\{\Pi_a,\mathbb{I}-\Pi_a\}$ associated with the PVM basis $\{\Pi_a\}$ of a finite-dimensional Hilbert space $\mathcal{H}$ is lower bounded as 
\begin{eqnarray}
\delta_{\{\Pi_a\}}(\varrho)&\ge&\frac{1}{2}\sup_{A\in\mathbb{H}(\mathcal{H}|\{\Pi_{a}\})}\sup_{B\in\mathbb{H}(\mathcal{H})}\big\{\big(\big|{\rm Tr}\{\varrho[\tilde{A}_{\varrho},\tilde{B}_{\varrho}]_-\}\big|^2\nonumber\\
&+&\big|{\rm Tr}\{\varrho[\tilde{A}_{\varrho},\tilde{B}_{\varrho}]_+\}-2{\rm Tr}\{\tilde{A}_{\varrho}\varrho\}{\rm Tr}\{\tilde{B}_{\varrho}\varrho\}\big|^2\big)^{1/2}\big\}-1. 
\label{lower bound for the state disturbance due to binary measurement}
\end{eqnarray}
\\
From Corollary 6a, we finally obtain the following trade-off relation, the proof of which follows similar steps to that of Proposition 5.\\
{\bf Corollary 6b}. Given a preparation represented by $\varrho$ on a finite-dimensional Hilbert space $\mathcal{H}$, the total disturbance $\delta_{\{\Pi_a\}}(\varrho)$ in the state $\varrho$ caused by the nonselective projective binary measurement $\{\Pi_a,\mathbb{I}-\Pi_a\}$ associated with the PVM basis $\{\Pi_a\}$, and the total disturbance $\delta_{\{\Pi_b\}}(\varrho)$ in the state $\varrho$ caused by the nonselective projective binary measurement $\{\Pi_b,\mathbb{I}-\Pi_b\}$ associated with the PVM basis $\{\Pi_b\}$, satisfy the following trade-off relation:
\begin{eqnarray}
\delta_{\{\Pi_a\}}(\varrho)\delta_{\{\Pi_b\}}(\varrho)&\ge&\frac{1}{4}\Big(\sup_{A\in\mathbb{H}(\mathcal{H}|\{\Pi_{a}\})}\sup_{B\in\mathbb{H}(\mathcal{H})}\big\{\big(\big|{\rm Tr}\{\varrho[\tilde{A}_{\varrho},\tilde{B}_{\varrho}]_-\}\big|^2\nonumber\\
&+&\big|{\rm Tr}\{\varrho[\tilde{A}_{\varrho},\tilde{B}_{\varrho}]_+\}-2{\rm Tr}\{\tilde{A}_{\varrho}\varrho\}{\rm Tr}\{\tilde{B}_{\varrho}\varrho\}\big|^2\big)^{1/2}\big\}-1\Big)^2.
\label{uncertainty relation for the state disturbance due to the nonselective binary measurement}
\end{eqnarray} 

\section{Summary and Discussion}

We have first derived lower bounds for the KD nonreality and the KD nonclassicality relative to a pair of rank-1 PVM bases, respectively in Eqs. (\ref{KD nonreality relative to a pair of PVMs is lower bounded by Robertson bound}) and (\ref{KD nonclassility relative to a pair of POVMs is lower bounded by Robertson-Schrodinger bound}). Nonvanishing lower bounds thus provide sufficient conditions for the KD quasiprobability to be nonclassical, i.e., its value is nonreal or its real part is negative, or both. We then defined the KD nonreality and KD nonclassicality in a state relative to a single PVM basis by taking the supremum over the other basis as in Eqs. (\ref{KD nonreality relative to a PVM}) and (\ref{KD nonclassicality relative to a PVM}). They can be interpreted as quantifying the amount of the quantumness in the state relative to the PVM basis manifesting their noncommutativity. We obtained lower bounds for the KD nonreality and KD nonclassicality in a state relative to a single PVM basis, given respectively in Eqs. (\ref{KD nonreality relative to a PVM is lower bounded by a normalized Robertson bound}) and (\ref{KD-nonclassility relative to a PVM is lower bounded by Robertson-Schrodinger bound 1}). A lower bound for the KD-nonreality in a state relative a rank-1 PVM in terms of extremal trace-norm asymmetry is given in Eq. (\ref{KD nonreality is lower bounded by a normalized trace-norm asymmetry}). The same lower bounds also apply to the corresponding $l_1$-norm coherence. 

We proceeded to derive trade-off relations for the KD nonreality and the KD nonclassicality relative to a PVM basis and those relative to another PVM basis  given in Eqs. (\ref{uncertainty relation for the KD nonreality relative to a PVM}) and (\ref{uncertainty relation for the KD classicality relative to a PVM}), having similar forms respectively to the Robertson and Roberston-Schr\"odinger uncertainty relations. The lower bounds for the trade-off relations involve optimization over two convex sets of Hermitian operators whose complete set of eigenprojectors are given by the corresponding PVM bases. We then showed that the trade-off relations imply similar trade-off relations for the $l_1$-norm coherence. The trade-off relations thus restrict simultaneous quantumness associated with a state $\varrho$ relative to two noncommuting rank-1 PVM basis. More detailed comparison of the uncertainty relations to the uncertainty relation for intrinsic quantum randomness presented in Refs. \cite{Korzekwa quantum-classical decomposition,Singh uncertainty relation for coherence,Yuan uncertainty relation for coherence,Hall quantum-classical decomposition} is left for future study.   

We further briefly discussed a hybrid quantum-classical variational scheme for a direct measurement of the KD nonreality and KD nonclassicality in a state relative a PVM basis by means of weak value measurement for the reconstruction of the KD quasiprobability, combined with a classical optimization scheme for searching the supremum over the set of rank-1 PVM bases of the Hilbert space. This operational interpretation leads to an information theoretical interpretation for the KD nonreality in a state relative a PVM basis as a lower bound for the total root-mean-squared error of the optimal estimation of the PVM basis based on the outcomes of projective measurement, in the worst case scenario. Moreover, it also leads to an uncertainty relation between the root-mean-squared error of the optimal estimation of the PVM basis and that of the optimal estimation of the other PVM basis, based on projective measurement, in the worst case scenario. We further applied the decomposition of the KD quasiprobability obtained via two successive projective measurements, into a real and nonnegative joint probability and two quantum modification terms which are responsible for the negativity and nonreality of the KD quasiprobability. Using this decomposition, the KD nonclassicality in a state relative to a PVM basis can be shown to give a lower bound to the total disturbance to the state caused by a nonselective projective binary measurement associated with the PVM basis. This further implies similar lower bound and trade-off relation for such total disturbance as those for the KD nonclassicality relative to a PVM basis.   

In this article, we have based all of our discussion on the standard KD quasiprobability associated with a density operator over a pair of rank-1 orthogonal PVM bases as in Eq. (\ref{standard KD quasiprobability}). This suggests directions for further investigation in the future. First, it is natural to ask if one can extend the methods and results of the present work to more general POVM (positive-operator-valued measure) bases. Next, recently, motivated by certain interesting physical problems such as quantum metrology with postselection \cite{Arvidsson-Shukur quantum advantage in postselected metrology,Lupu-Gladstein negativity enhanced quantum phase estimation 2022} and detection of OTOC (out-of-time-order correlation) in many body chaotic system \cite{Halpern quasiprobability and information scrambling,Alonso KD quasiprobability witnesses quantum scrambling}, there are proposals to extend the KD quasiprobability by extending the number of PVM basis. Within the representation of the KD quasiprobability via weak value, this extension means that we increase the number of weak measurements before making strong projective postselection measurement. This extension of the KD quasiprobability too shares the properties of the standard KD quasiprobability. In particular, its negativity and nonreality signal quantumness associated with quantum noncommutativity. It is therefore interesting to apply the methods and reasoning developed in the present article and also in Refs. \cite{Agung KD-nonreality coherence,Agung KD-nonclassicality coherence,Agung KD general quantum correlation} to use the extended KD quasiprobability to probe quantum coherence, general quantum correlation and to see the restriction imposed by the uncertainty principle. Such an approach might in turn help clarify the roles of coherence and general correlation in quantum metrology with postselection and OTOC. 

\begin{acknowledgments} 
\end{acknowledgments} 

\appendix 

\section{Proof of Proposition 2 \label{A proof of Proposition 2}} 

Notice that the left-hand side of Eq. (\ref{KD nonreality relative to a PVM is lower bounded by a normalized Robertson bound}) does not depend on $\tilde{B}=B/\|B\|_{\infty}$. Hence, the inequality in Eq. (\ref{KD nonreality relative to a PVM is lower bounded by a normalized Robertson bound}) can be strengthened to obtain 
\begin{eqnarray}
\mathcal{Q}_{\rm KD}^{\rm NRe}(\varrho;\{\Pi_{a}\})&\ge&\frac{1}{2}\sup_{A\in\mathbb{H}(\mathcal{H}|\{\Pi_{a}\})}\sup_{B\in\mathbb{O}(\mathcal{H})}\big|{\rm Tr}\{\tilde{B}[\tilde{A},\varrho]_-\}\big|,
\label{KD nonreality is lower bounded by a normalized trace-norm asymmetry step 1}
\end{eqnarray}
where $\mathbb{O}(\mathcal{H})$ is the set of all bounded operators on $\mathcal{H}$. Next, since $\|\tilde{B}\|_{\infty}=1$, then one can further strengthen the inequality in Eq. (\ref{KD nonreality is lower bounded by a normalized trace-norm asymmetry step 1}) as 
\begin{eqnarray}
\mathcal{Q}_{\rm KD}^{\rm NRe}(\varrho;\{\Pi_{a}\})\ge\frac{1}{2}\sup_{A\in\mathbb{H}(\mathcal{H}|\{\Pi_{a}\})}\big\{\sup_{X\in\mathbb{O}(\mathcal{H}|\|X\|_{\infty}\le 1)}\big|{\rm Tr}\{X^{\dagger}[\tilde{A},\varrho]_-\}\big|\big\}, 
\label{KD nonreality is lower bounded by a normalized trace-norm asymmetry step 2}
\end{eqnarray}
where $\mathbb{O}(\mathcal{H}|\|X\|_{\infty}\le 1)$ is the set of all bounded operators $X$ with $\|X\|_{\infty}\le 1$. One then observes that the term on the right-hand side of Eq. (\ref{KD nonreality is lower bounded by a normalized trace-norm asymmetry step 2}) inside the bracket $\{\dots\}$ is just the variational definition of the Schatten $p=1$ norm via its conjugate norm $p_*=\infty$ \cite{Watrous book on quantum Shannon theory}, so that one has 
\begin{eqnarray}
\sup_{X\in\mathbb{O}(\mathcal{H}|\|X\|_{\infty}\le 1)}\big|{\rm Tr}\{X^{\dagger}[\tilde{A},\varrho]_-\}\big|=\|[\tilde{A},\varrho]_-\|_1. 
\end{eqnarray}
Inserting this into the right-hand side of Eq. (\ref{KD nonreality is lower bounded by a normalized trace-norm asymmetry step 2}), we obtain Eq. (\ref{KD nonreality is lower bounded by a normalized trace-norm asymmetry}). 

As an alternative scheme of proof, first, from Proposition 4 of Ref. \cite{Agung estimation and operational interpretation of trace-norm asymmetry}, we have (see Eq. (22) of Ref. \cite{Agung estimation and operational interpretation of trace-norm asymmetry})
\begin{eqnarray}
\mathcal{Q}_{\rm KD}^{\rm NRe}(\varrho;\{\Pi_{a}\})\ge\|[A,\varrho]_-\|_1/2\|A\|_{\infty}, 
\label{proof of l1-norm coherence is lower bounded by maximum trace norm asymmetry appendix step 1}
\end{eqnarray}
where $\mathcal{Q}_{\rm KD}^{\rm NRe}(\varrho;\{\Pi_{a}\})$ in this article is denoted by $C_{\rm KD}(\varrho;\{\ket{a}\})$ in Ref. \cite{Agung estimation and operational interpretation of trace-norm asymmetry}. This can be proven by using the equality between the trace-norm asymmetry \cite{Marvian - Spekkens speakable and unspeakable coherence} and the measure of asymmetry based on the maximum mean absolute value of the imaginary part of the weak value of the generator $A$ of the translation group proposed in \cite{Agung translational asymmetry from nonreal weak value} as expressed in Proposition 1 of Ref. \cite{Agung estimation and operational interpretation of trace-norm asymmetry} (see Eq. (5) of Ref. \cite{Agung estimation and operational interpretation of trace-norm asymmetry}). Observe that the left-hand side of Eq. (\ref{proof of l1-norm coherence is lower bounded by maximum trace norm asymmetry appendix step 1}) is independent of the eigenvalues spectrum of the Hermitian operator $A$ which appears on the right-hand side. Hence, the inequality can be strengthened as
\begin{eqnarray}
\mathcal{Q}_{\rm KD}^{\rm NRe}(\varrho;\{\Pi_{a}\})\ge\sup_{A\in\mathbb{O}(\mathcal{H}|\{\Pi_{a}\})}\|[A,\varrho]_-\|_1/2\|A\|_{\infty}. 
\end{eqnarray} 
\qed

\section{Propositions 1 and 2 for two-dimensional Hilbert space \label{Proof that the equality in lemma 1 is achieved for a single qubit}}

First, assume without loss of generality, that the PVM basis of the two-dimensional Hilbert space $\mathcal{H}\cong\mathbb{C}^2$ relative to which we define the KD nonreality in a state on the left-hand side of Eq. (\ref{KD nonreality relative to a PVM is lower bounded by a normalized Robertson bound}) is given by the complete set of eigenprojectors of the Pauli operator $\sigma_z$, i.e., $\mathbb{A}:=\{\ket{0}\bra{0},\ket{1}\bra{1}\}$. All the Hermitian operators on $\mathbb{C}^2$ with the eigenprojectors $\mathbb{A}$ thus take the general form as:
\begin{eqnarray}
A=a_0\ket{0}\bra{0}+a_1\ket{1}\bra{1},
\label{Hermitian operators with eigenprojectors given by the eigenprojectors of Pauli z} 
\end{eqnarray}
where $a_0,a_1\in\mathbb{R}$ are the eigenvalues. We denote the set of all such Hermitian operators by $\mathbb{H}(\mathbb{C}^2|\mathbb{A})$. Moreover, the general form of all Hermitian operators on the Hilbert space $\mathbb{C}^2$ reads as    
\begin{eqnarray}
&&B(\alpha,\beta)=b_+\ket{b_+(\alpha,\beta)}\bra{b_+(\alpha,\beta)}+b_-\ket{b_-(\alpha,\beta)}\bra{b_-(\alpha,\beta)},
\label{general Hermitian operator on two-dimensional Hilbert space}
\end{eqnarray} 
with the eigenvalues $b_+,b_-\in\mathbb{R}$, and the corresponding orthonormal eigenvectors $\{\ket{b_+(\alpha,\beta)},\ket{b_-(\alpha,\beta)}\}$ can be expressed using the Bloch sphere parameterization as 
\begin{eqnarray}
\ket{b_+(\alpha,\beta)}&=&\cos\frac{\alpha}{2}\ket{0}+e^{i\beta}\sin\frac{\alpha}{2}\ket{1};\nonumber\\
\ket{b_-(\alpha,\beta)}&=&\sin\frac{\alpha}{2}\ket{0}-e^{i\beta}\cos\frac{\alpha}{2}\ket{1}, 
\label{complete set of basis in the x-y plane}
\end{eqnarray}
where $\alpha\in[0,\pi]$, $\beta\in[0,2\pi)$. Let us denote the set of all Hermitian operators on $\mathbb{C}^2$ by $\mathbb{H}(\mathbb{C}^2)$. 

We further assume, without loss of generality, that the singular values of $A$ and $B$ have the following orderings: $|a_0|\ge|a_1|$ and $|b_+|\ge|b_-|$, so that we have $\|A\|_{\infty}=|a_0|$ and $\|B\|_{\infty}=|b_+|$. Then, computing the lower bound in Eq. (\ref{KD nonreality relative to a PVM is lower bounded by a normalized Robertson bound}), we obtain 
\begin{eqnarray}
&&\mathcal{Q}_{\rm KD}^{\rm NRe}(\varrho;\mathbb{A})\nonumber\\
\label{lower bound for KD nonreality relative to PVM for a single qubit is equal to the l1-norm coherence step 1}
&\ge&\sup_{A\in\mathbb{H}(\mathbb{C}^2|\mathbb{A})}\sup_{B\in\mathbb{H}(\mathbb{C}^2)}\frac{|{\rm Tr}\{B[A,\varrho]_-\}|}{2\|A\|_{\infty}\|B\|_{\infty}}\\
\label{lower bound for KD nonreality relative to PVM for a single qubit is equal to the l1-norm coherence step 2}
&=&\frac{1}{2}\max_{\{a_0,a_1\}\in\mathbb{R}^2}\max_{\{b_+,b_-\}\in\mathbb{R}^2}\max_{\{\alpha,\beta\}\in[0,\pi]\times[0,2\pi)}\Big\{|\sin\alpha\sin(\beta-\phi_{01})|\nonumber\\
&\times&\frac{|b_+-b_-|}{|b_+|}\frac{|a_0-a_1|}{|a_0|}|\braket{1|\varrho|0}|\Big\}\\
&=&\frac{1}{2}\max_{\{a_0,a_1\}\in\mathbb{R}^2}\max_{\{b_+,b_-\}\in\mathbb{R}^2}\Big\{\frac{|b_+-b_-|}{|b_+|}\frac{|a_0-a_1|}{|a_0|}|\braket{0|\varrho|1}|\Big\}\\
\label{lower bound for KD nonreality relative to PVM for a single qubit is equal to the l1-norm coherence step 4}
&=&2|\braket{0|\varrho|1}|=C_{l_1}(\varrho;\mathbb{A}). 
\end{eqnarray} 
Here, $\phi_{01}=-\arg\braket{0|\varrho|1}$, the maximum is obtained for Hermitian operator $A$ of Eq. (\ref{Hermitian operators with eigenprojectors given by the eigenprojectors of Pauli z}) with $|a_0-a_1|=2|a_0|$ and for Hermitian operator $B(\alpha,\beta)$ having the form of (\ref{general Hermitian operator on two-dimensional Hilbert space}) with $\alpha=\pi/2$ and $\beta=\phi_{01}+\pi/2$ and $|b_+-b_-|=2|b_+|$, and $C_{l_1}(\varrho;\mathbb{A})=2|\braket{0|\varrho|1}|$ is the $l_1$-norm coherence of $\varrho$ relative to the orthonormal basis $\mathbb{A}=\{\ket{0},\ket{1}\}$. Note that, to get Eq. (\ref{lower bound for KD nonreality relative to PVM for a single qubit is equal to the l1-norm coherence step 4}), we have used the fact that for any pair $x,y\in\mathbb{C}$ with $|x|\ge|y|$, we always have $|x-y|\le 2|x|$, and equality is attained for $x=-y$. On the other hand, as shown in Ref. \cite{Agung KD-nonreality coherence}, for arbitrary state of a single qubit and any PVM basis of $\mathbb{C}^2$, the left-hand side of Eq. (\ref{lower bound for KD nonreality relative to PVM for a single qubit is equal to the l1-norm coherence step 1}) is also given by the $l_1$-norm coherence, i.e.:
\begin{eqnarray}
\mathcal{Q}_{\rm KD}^{\rm NRe}(\varrho;\mathbb{A})=2|\braket{0|\varrho|1}|=C_{l_1}(\varrho;\mathbb{A}).
\label{equality between KD nonreality and l1-norm coherence for a single qubit}
\end{eqnarray} 
Hence, the inequality in Eq. (\ref{lower bound for KD nonreality relative to PVM for a single qubit is equal to the l1-norm coherence step 1}) indeed becomes equality. Moreover, one can see from the values of $\alpha$ and $\beta$ that achieve the maximum in Eq. (\ref{lower bound for KD nonreality relative to PVM for a single qubit is equal to the l1-norm coherence step 2}), that the eigenbasis of $B_*$ expressed in Eq. (\ref{complete set of basis in the x-y plane}) which attains the supremum in Eq. (\ref{lower bound for KD nonreality relative to PVM for a single qubit is equal to the l1-norm coherence step 1}) are mutually unbiased with $\mathbb{A}=\{\ket{0},\ket{1}\}$ and also with the eigenbasis of $\varrho$. This proves Proposition 1 for the case of two-dimensional Hilbert space.

Next, one can see from the proof of Proposition 2 in Appendix \ref{A proof of Proposition 2} that the lower bound in Eq. (\ref{KD nonreality relative to a PVM is lower bounded by a normalized noncommutativity}) is less than or equal to the lower bound in Eq. (\ref{KD nonreality is lower bounded by a normalized trace-norm asymmetry}), and the left-hand sides of the two equations are the same. Hence, when the inequality in Eq. (\ref{KD nonreality relative to a PVM is lower bounded by a normalized noncommutativity}) becomes equality, the inequality in Eq. (\ref{KD nonreality is lower bounded by a normalized trace-norm asymmetry}) must also become equality. This combined with the above result means that for two-dimensional Hilbert space, the inequality in Eq. (\ref{KD nonreality is lower bounded by a normalized trace-norm asymmetry}) must also become equality. Indeed, computing the trace-norm asymmetry of the state $\varrho$ relative to the translation group generated by $A$ having the form of Eq. (\ref{Hermitian operators with eigenprojectors given by the eigenprojectors of Pauli z}), one has $\|[A,\varrho]\|_1/2=|a_0-a_1||\braket{0|\varrho|1}|$. Upon inserting into the lower bound in Eq. (\ref{KD nonreality is lower bounded by a normalized trace-norm asymmetry}), we have 
\begin{eqnarray}
&&\sup_{A\in\mathbb{H}(\mathcal{H}|\{\Pi_{a}\})}\|[A,\varrho]_-\|_1/2\|A\|_{\infty}\nonumber\\
&=&\max_{\{a_0,a_1\}\in\mathbb{R}^2}\frac{|a_0-a_1|}{|a_0|}|\braket{0|\varrho|1}|\nonumber\\
&=&2|\braket{0|\varrho|1}|\nonumber\\
&=&\mathcal{Q}_{\rm KD}^{\rm NRe}(\varrho;\mathbb{A}),
\label{Proposition 2 for two-dimensional Hilbert space}
\end{eqnarray}
where the last equality in Eq. (\ref{Proposition 2 for two-dimensional Hilbert space}) is just Eq. (\ref{equality between KD nonreality and l1-norm coherence for a single qubit}) and the maximum is obtained when $|a_0-a_1|=2|a_0|$. This proves Proposition 2 for the case of two-dimensional Hilbert space.

Combining Eqs. (\ref{equality between KD nonreality and l1-norm coherence for a single qubit}) and (\ref{Proposition 2 for two-dimensional Hilbert space}), we thus obtain Eq. (\ref{KD-nonreality for two-dimensional Hilbert space}) of the main text. 
\qed

\section{Trade-off relation of Eq. (\ref{additive uncertainty relation for l1-norm coherence}) for a single qubit\label{Additive l1 norm trade-off relation for a single qubit}} 

Without loss of generality, one can proceed as in Appendix \ref{Proof that the equality in lemma 1 is achieved for a single qubit}, but now the optimization is over the set $\mathbb{H}(\mathbb{C}^2|\mathbb{A})$ of all Hermitian operators on $\mathbb{C}^2$ with the complete set of eigenprojectors $\mathbb{A}=\{\ket{0}\bra{0},\ket{1}\bra{1}\}$ having the form of Eq. (\ref{Hermitian operators with eigenprojectors given by the eigenprojectors of Pauli z}), and over the set $\mathbb{H}(\mathbb{C}^2|\mathbb{B})$ of all Hermitian operators on $\mathbb{C}^2$ with the complete set of eigenprojectors $\mathbb{B}=\{\ket{b_+(\alpha,\beta)}\bra{b_+(\alpha,\beta)},\ket{b_-(\alpha,\beta)}\bra{b_+(\alpha,\beta)}\}$ having the form of Eq. (\ref{general Hermitian operator on two-dimensional Hilbert space}). Evaluating the lower bound in Eq. (\ref{additive uncertainty relation for l1-norm coherence}) for two-dimensional Hilbert space, we have 
\begin{eqnarray}
&&C_{l_1}(\varrho;\mathbb{A})+C_{l_1}(\varrho;\mathbb{B})\nonumber\\
&\ge&\sup_{A\in\mathbb{H}(\mathbb{C}^2|\mathbb{A})}\sup_{B(\alpha,\beta)\in\mathbb{H}(\mathbb{C}^2|\mathbb{B})}\frac{|{\rm Tr}\{[A,B(\alpha,\beta)]\varrho\}|}{\|A\|_{\infty}\|B\|_{\infty}}\nonumber\\
&=&\max_{\{b_+,b_-\}\in\mathbb{R}^2}\max_{\{a_0,a_1\}\in\mathbb{R}^2}\frac{|a_0-a_1|}{|a_0|}\frac{|b_+-b_-|}{|b_+|}|\braket{0|\varrho|1}|\nonumber\\
&\times&|\sin\alpha\sin(\beta-\phi_{01})|\nonumber\\
&=&4|\braket{0|\varrho|1}||\sin\alpha||\sin(\beta-\phi_{01})|\nonumber\\
&=&2\sqrt{r^2-r_z^2}|\sin\alpha||\sin(\beta-\phi_{01})|\nonumber\\
&=&2r|\sin\phi_z||\sin\alpha||\sin(\beta-\phi_{01})|.
\label{lower bound for UR for KD nonreality relative to PVMs for a single qubit 0}
\end{eqnarray} 
Here, the maximum is obtained when $|a_0-a_1|=2|a_0|$ and $|b_+-b_-|=2|b_+|$ (see Appendix \ref{Proof that the equality in lemma 1 is achieved for a single qubit}), we have used the expression for the qubit state $\varrho=\frac{1}{2}(\mathbb{I}+r_x\sigma_x+r_y\sigma_y+r_z\sigma_z)$, $r_x^2+r_y^2+r_z^2=r^2$ so that $2|\braket{0|\varrho|1}|=|r_x-ir_y|=\sqrt{r^2-r_z^2}$, and $\phi_z$ is the angle between the Bloch vector of the state and the positive $z$-axis. One can see that the lower bound decreases as the purity of the state given by ${\rm Tr}(\varrho^2)=(1+r^2)/2$ decreases. Moreover, it also decreases when the noncommutativity between the two PVM bases, i.e., $\mathbb{A}$ and $\mathbb{B}$, quantified by $|\sin\alpha|$, decreases. In particular, the lower bound vanishes for $r=0$, i.e., for maximally mixed state  $\varrho=\mathbb{I}/2$ with minimum purity, and it also vanishes when $\sin\alpha=0,\pi$, i.e., when the two PVM bases $\mathbb{A}$ and $\mathbb{B}$ are commuting. This result is in accord with that obtained in Ref. \cite{Yuan uncertainty relation for coherence}. Note that $|\sin\phi_z|$ and $|\sin(\beta-\phi_{01})|$ on the right-hand side characterize respectively the noncommutativity between the state $\varrho$ and the PVM basis $\mathbb{A}$ and between $\varrho$ and the PVM basis $\mathbb{B}$. They vanish respectively when $\varrho$ commutes with $\mathbb{A}$ and $\varrho$ commutes with $\mathbb{B}$, as expected. As an example consider the case when the state is pure so that $r=1$, and take $\alpha=\pi/2$ and $\phi_{01}+\pi/2=\beta$ so that $\sin\alpha=\sin(\beta-\phi_{01})=1$. Then, taking $\phi_z=\pi/2$, we have $C_{l_1}(\varrho;\mathbb{A})=C_{l_1}(\varrho;\mathbb{B})=1$ and the inequality in Eq. (\ref{lower bound for UR for KD nonreality relative to PVMs for a single qubit 0}) becomes equality, i.e., both sides are equal to $2$. Note that in this case, the triple $\mathbb{A}$, $\mathbb{B}$ and the eigenbasis of $\varrho$ comprise a three mutually unbiased bases of $\mathbb{C}^2$.

\section{Proof of Eq. (\ref{l1-norm coherence is lower bounded by the KD nonclassicality relative to a PVM}) \label{Proof that KD nonclassicality coherence gives a lower bound to the l1-norm coherence}} 

One first has, from the definition of the KD nonclassicality in a state $\varrho$ relative to a PVM $\{\Pi_a\}$ in Eq. (\ref{KD nonclassicality relative to a PVM}), 
\begin{eqnarray}
&&\mathcal{Q}_{\rm KD}^{\rm NCl}(\varrho;\{\Pi_a\})\nonumber\\
&=&\sup_{\{\Pi_{b}\}\in\mathcal{M}_{\rm r1PVM}(\mathcal{H})}\sum_{a,b}\big|\sum_{a'}{\rm Tr}\{\Pi_b\Pi_a\varrho\Pi_{a'}\}\big|-1\nonumber\\
&\le&\sum_{a,a'}\big|\braket{a|\varrho|a'}|\sum_{b_*}|\braket{a'|b_*}\braket{b_*|a}\big|-1, 
\label{KD-nonclassicality coherence is upper bounded by l1 coherence step 0}
\end{eqnarray}
where we have used a completeness relation $\sum_{a'}\Pi_{a'}=\mathbb{I}$, triangle inequality, and $\{\Pi_{b_*}\}\in\mathcal{M}_{\rm r1PVM}(\mathcal{H})$ is a PVM basis which achieves the supremum. On the other hand, using the Cauchy-Schwartz inequality, we have $\sum_{b_*}|\braket{b_*|a}\braket{a'|b_*}|\le(\sum_{b_*}|\braket{b_*|a}|^2\sum_{b_*'}|\braket{a'|b'_*}|^2)^{1/2}=1$, where we have used a completeness relation $\sum_{b_*}\ket{b_*}\bra{b_*}=\mathbb{I}$. Inserting this into Eq. (\ref{KD-nonclassicality coherence is upper bounded by l1 coherence step 0}), we finally obtain
\begin{eqnarray}
\mathcal{Q}_{\rm KD}^{\rm NCl}(\varrho;\{\Pi_a\})&\le&\sum_{a,a'}\big|\braket{a|\varrho|a'}|-1\nonumber\\
&=&\sum_{a\neq a'}\big|\braket{a|\varrho|a'}|\nonumber\\
&=&C_{l_1}(\varrho;\{\ket{a}\}),
\label{KD-nonclassicality coherence is upper bounded by l1 coherence}
\end{eqnarray} 
where we have used the normalization $\sum_a\braket{a|\varrho|a}=1$. \qed

\end{document}